\documentclass[journal]{IEEEtran}

\usepackage{cite,graphicx,amsmath,amssymb}
\usepackage{subfigure}
\usepackage{fancyhdr}
\usepackage{mdwmath}
\usepackage{mdwtab}
\usepackage{balance}
\usepackage{xcolor}
\usepackage{bm}
\usepackage{amsthm}
\usepackage{algorithm}
\usepackage{algorithmic}
\usepackage{multirow}
\usepackage{flafter}

\newtheorem{lemma}{Lemma}
\newtheorem{theorem}{Theorem}
\newtheorem{remark}{Remark}
\newtheorem{corollary}{Corollary}
\newtheorem{proposition}{Proposition}

\begin{document}

\title{Performance Analysis of Clustered LoRa Networks}
\author{
\IEEEauthorblockN{Zhijin~Qin,~\IEEEmembership{Member,~IEEE,}
 Yuanwei~Liu,~\IEEEmembership{Senior Member,~IEEE,}
Geoffrey~Ye~Li,~\IEEEmembership{Fellow,~IEEE,} and~Julie~A.~McCann,~\IEEEmembership{Member,~IEEE.}
 }
 \thanks{Part of the work has been presented in IEEE International Conference on Communications, Paris, France, May, 2017~\cite{Zhijin:ICC:2017}. This work was funded by ICRI.}
\thanks{Zhijin Qin and Yuanwei Liu are with Queen Mary University of London, London E1 4NS, UK, (e-mail: \{z.qin; yuanwei.liu\}@qmul.ac.uk). Zhijin Qin is also with Imperial College London as an Honorary Research Fellow.}
\thanks{Geoffrey Ye Li is with Georgia Institute of Technology, Atlanta, GA, USA, 30332-0250 (e-mail: liye@ece.gatech.edu).}
\thanks{Julie A. McCann is with Imperial College London, London SW7 2AZ, UK, (email: j.mccann@imperial.ac.uk).}
}

\maketitle

\begin{abstract}
In this paper, we investigate the uplink transmission performance of low-power wide-area (LPWA)  networks with regards to coexisting radio modules. We adopt long range (LoRa) radio technique as an example of the network of focus even though our analysis can be easily extended to other situations. We exploit a new topology to model the network, where the node locations of LoRa follow a Poisson cluster process (PCP) while other coexisting radio modules follow a Poisson point process (PPP). Unlike most of the performance analysis based on stochastic geometry, we take noise into consideration. More specifically, two models, with a fixed  and a random number of active LoRa nodes in each cluster, respectively, are considered. To obtain insights, both the exact and simple approximated expressions for coverage probability are derived. Based on them, area spectral efficiency and energy efficiency are obtained. From our analysis, we show how the performance of LPWA networks  can be enhanced through adjusting the density of LoRa nodes around each LoRa receiver. Moreover, the simulation results unveil that the optimal number of active LoRa nodes in each cluster exists to maximize the area spectral efficiency.
\end{abstract}


\section{Introduction}
Internet of Things (IoT) is envisioned as a means to connect billions of small computing devices embedded in different environments (e.g., walls and soil) and even implanted in human bodies~\cite{KIM:2016,Xie2017,QIN:SPM:2018,Tao:TII:2018}. Aiming to provide  connectivity opportunities for massive numbers of devices, two possible networking approaches have been proposed. One is the evolution from the existing communication systems, i.e., fifth generation (5G) or the beyond for intelligent communications~\cite{Qin:2019},  with the purpose of supporting machine-type communications (MTC)~\cite{sharma2018towards,Li:WM:2018}. Another is to design MTC-dedicated networks from scratch, such as low-power wide-area  (LPWA) networks~\cite{Xiong_magazine:2015,Mischa_JSAC:2016,IEEE_WIRELESS:2016,QIN:CM:2018}.

In LPWA networks, the transmission range is a dedicating factor for highly scalable smart metering or other related applications where only a small portion of data is transmitted, perhaps after considerable analysis or filtering from sensors. Another critical factor in LPWA  networks  is energy consumption since they consist of energy-constrained devices. The battery lifetime of some smart city sensors is required to be no less than ten years for IoT applications~\cite{LPWA}. Two technical options have been proposed for LPWA networks to enhance  signal-to-noise ratio (SNR) ratio and to increase transmission ranges with enhanced power efficiency; they are the ultra narrrowband approach and the coding gain approach. The ultra narrowband approach enhances SNR by focusing signal in a narrowband. One of the classic technologies, Sigfox~\cite{Sigfox} uses the narrowband approach to implement an LPWA network. The other approach exploits coding gain to combat high noise power in a wideband receiver. Long range (LoRa) radio technique~\cite{LoRa_specification} is an example of the code gain approach.

LoRa has a relatively long transmission range and low energy consumption, which has attracted much attention in IoT field and became the most widely deployed LPWA technique. Furthermore, the chirp spread spectrum technology used in LoRa allows the usage of cheap oscillators with high stability guaranteed at the receiver~\cite{LoRa_specification}. These advantages make LoRa a popular candidate for smart city scenarios. LoRaWAN is a medium access control (MAC) protocol for LoRa, and supports star-topologies  providing high capacity and longer transmission ranges~\cite{LoRa_specification}. The uplink of LoRaWAN is scheduled by end-devices based on their transmission requirements, regardless of the channel occupancy. LoRaWAN achieves low energy consumption since it does not sense the medium before sending its packet and does not require any synchronization to access the medium. However, packet collision remains a problem. The coverage probability of  LoRa networks has been analyzed~\cite{Georgiou:2017} by considering the interfering signals using the same spreading factor (SF). However, we believe that both intra-interference, i.e., from LoRa users in the same and neighbouring networks, and inter-interference, i.e., from other LPWA users sharing the same spectrum, combat the performance of LPWA networks~\cite{Zhijin:ICC:2017}. So far, some experiments on LoRa and other LPWA techniques have been carried out as the initial trials in wireless sensor networks~\cite{Bor:2016:LIT,Bor:2016:LLW,Deng2017,Voigt:2017,QIN:CM:2018,Froytlog:2019}. Particularly, we have deployed LoRa devices in the Queen Elizabeth Park, London, to collect data  and installed a LoRa gateway to forward the collected data to a cloud server for further processing and analyzing~\cite{QIN:CM:2018}. Based on this implementation, we have further investigated the resource efficiency and energy efficiency in LoRa networks by optimizing the channel selection and transmit power of LoRa users~\cite{QIN:GC:2017,QIN:GC:2018}.

To make this work more general,  we theoretically analyze the performance of clustered LPWA networks  with particular focus on the interference from coexisting LoRa users and other LPWA   users, as they all work over unlicensed spectrum. Stochastic geometry is a powerful mathematical tool for designing and analyzing wireless networks, particularly dense networks~\cite{Martin_survey:2009}, such as LPWA networks  that potentially provide massive connectivity. When wireless nodes are uniformly distributed in an area, homogeneous Poisson point processes (PPPs) can accurately model dense networks. However, when sensors are clustered, such as in smart city scenarios, PPP is unable to model interference~\cite{Martin_TIF_2009}. In this situation, Poisson cluster process (PCP), where parent points form a PPP and offspring points form clusters centred at the parent point, is necessary to model wireless networks using random cluster topologies arising from geographical factors or MAC protocols~\cite{Martin_TIF_2009,Gulati_TSP:2010}.

PCPs have attracted lots of attention in cellular networks and wireless sensor networks~\cite{Heath2013Tcom,Chun_JSAC:2015,Vinay_TWC:2015,Kaibing:2016,Afshang_TWC:2016,Tang:2017,Tang:TVT:2018}, recently. In \cite{Heath2013Tcom}, an interference alignment approach has been used for a cluster topology to address intra-cluster interference for multiple-input multiple-output (MIMO) systems, where a spatial PCP process is used to model the nodes in a random access network.  For heterogeneous cellular networks, PCP has been adopted in~\cite{Chun_JSAC:2015,Vinay_TWC:2015} to model nodes clustering at hot spots while taking into consideration the fact that base stations belonging to different tiers may differ in terms of transmit power, node density, and link reliability. In~\cite{Kaibing:2016}, the PCP model has been applied to wirelessly powered backscatter communications, where power beacons (PBs) form the parent PPP and the backscatter nodes are children points in the cluster. A \emph{Thomas} PCP model can capture the fact that a given device typically has multiple proximate devices, any of which can potentially act as a serving device. It has been used to model the device locations for device-to-device (D2D) networks in~\cite{Afshang_TWC:2016}. Moreover, a \emph{Matern} PCP has been used to model wireless networks exhibiting device clustering and the distance distribution has been derived to describe the interference statistics and connection probability in clustered networks~\cite{Tang:2017}. Based on the results, the \emph{Matern} PCP and modified Thomas PCP have been investigated when they are adopted to model the out-of-band D2D networks~\cite{Tang:TVT:2018}.  Furthermore, it has been pointed out that clustering is beneficial for long range transmissions in ad-hoc networks~\cite{Martin_TIF_2009}. Therefore, the PCP can model LPWA networks well.

While the aforementioned research contributions have laid a solid foundation and provided a good understanding on the PCP model, the performance analysis using PPP/PCP based stochastic geometry approach to investigate LPWA networks  is still missing. Different from the clustered D2D networks, LoRa works over unlicensed spectrum, which makes the network suffering from the most severe interference caused by: i) The LoRa nodes in the same cluster, which is due to the non-orthogonality of SFs; ii). The LoRa nodes clustered in the neighbouring clusters accessing the same channel; iii). The non-LoRa nodes that are using other radio access networks over the same unlicensed channel. In this paper, we adopt \emph{Matern} cluster process~\cite{Kroese2012SpatialPG} to model LPWA networks, where the LoRa receivers forms PPP cluster centers and the active LoRa nodes in each cluster form the children process  since sensor nodes, like metering sensors distributed in a building, are highly clustered  based on geography. We also assume that the coexisting non-LoRa nodes are modelled as PPP. This topology is motivated by the fact that we can only control the clustering deployment of LoRa nodes rather than non-LoRa ones. Therefore, non-LoRa nodes may use any coexisting radio  and locate at any location within the considered area. Moreover, unlike existing research that mostly considers the interference-limited case, we additionally consider the impact of noise on the system performance. The reason  is  that LPWA networks  are not necessary interference-limited due to the long transmission distance with relatively low transmit power.

To the best of our knowledge, this work is the first attempt to model and analyze LPWA networks  using the \emph{Matern} PCP model. In this paper, we attempt to explore the potential performance enhancement brought by PCP as well as to answer the following questions:

\begin{enumerate}
  \item What is the impact in terms of the cluster radius on the system performance?
  \item Is there an optimal number of active LoRa nodes in each cluster?
  \item What is the most appropriate transmit power for LoRa users?
\end{enumerate}

The major contributions of this paper are summarized as follows:

\begin{enumerate}
  \item We use the spatial distributions of PCP to model the LoRa system by considering the potential interference from both co-existing LoRa users and non-LoRa users working over the same channel. Specifically, By considering the uplink transmission of LoRa networks, the \emph{Marten} cluster process is invoked to model the locations of LoRa nodes as well as receivers. Moreover, the PPP model is adopted to model non-LoRa nodes communicating over the same channel as the LoRa users.
  \item To characterize the performance of the LPWA networks, three performance metrics, including coverage probability, area spectral efficiency, and energy efficiency, are adopted. By using Gaussian-Chebyshev approximation, we derive both exact and simple approximated expressions of these three metrics with considering the impact of channel noise.
  \item We analytically prove that significant performance gains can be achieved by decreasing the radius of each cluster. We also demonstrate that there exists an optimal number of active LoRa nodes for each cluster that can maximize the area spectral efficiency. These remarks provide insightful guidelines for the implementation of LPWA networks.
\end{enumerate}

The rest of the paper is organized as follows. In Section II, the system model for LPWA networks  with LoRa is introduced. In Section III, new analytical expressions  for the coverage probability of the considered networks are derived. Then area spectral efficiency and energy efficiency are investigated in Section IV. Numerical results are presented in Section V, which is followed by the conclusions in Section VI.

\section{System Model}
In this section, the system settings and propagation model are introduced for the considered LoRa LPWA networks   and other coexisting LPWA radio modules that work over the same frequency.
\subsection{Spatial Setup and Key Assumptions}
\begin{figure}[t!]
    \begin{center}
        \includegraphics[width=3.2in,height=2.2in]{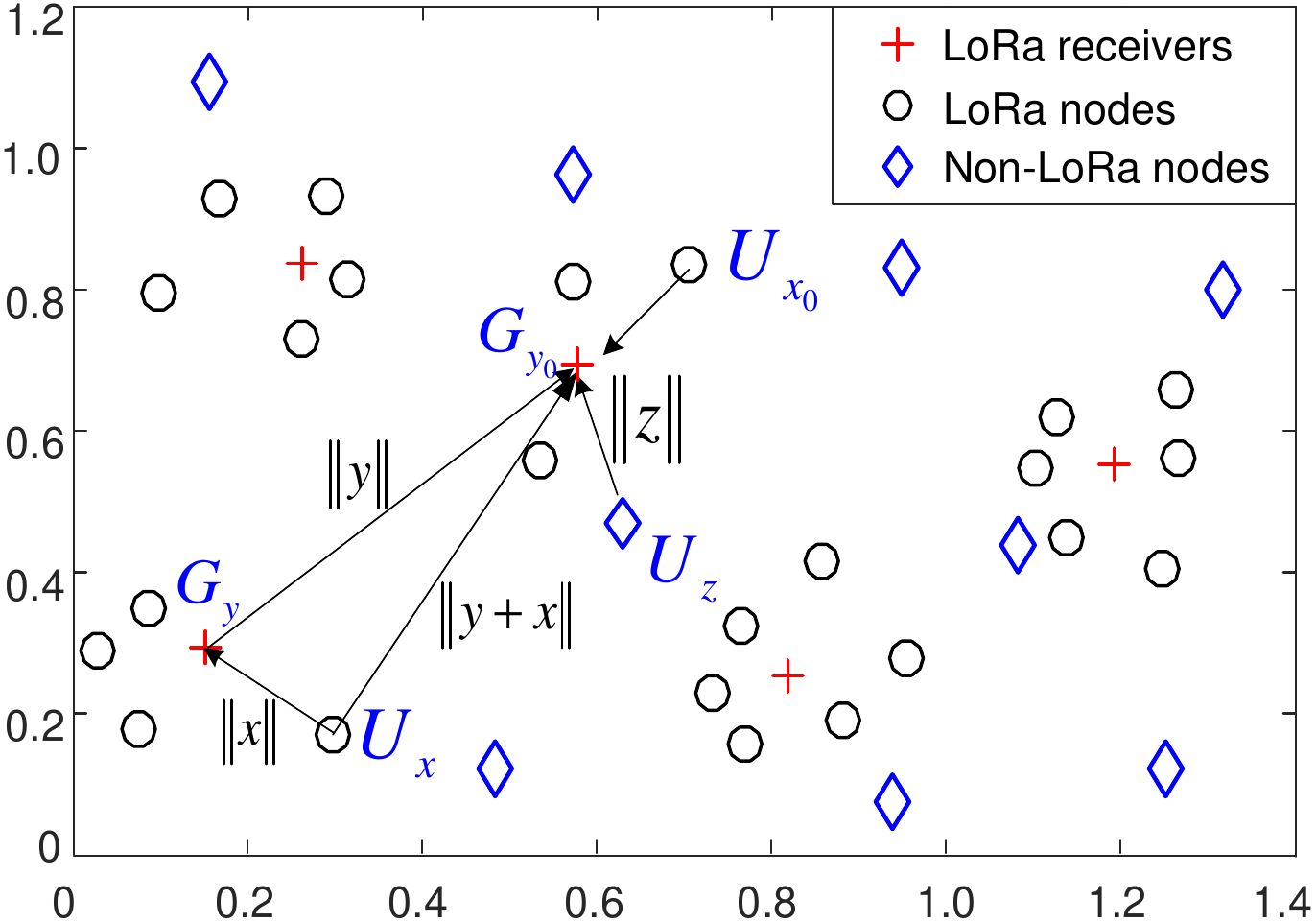}
        \caption{System model of the clustered low-power wide-area networks with LoRa and coexisting radios.}
        \label{system_model}
    \end{center}
\end{figure}

We consider the uplink transmission in LPWA networks  where the system under study is LoRa with interference from other LPWA radio modules (e.g., Sigfox). A star-topology is adopted and the nodes use single-hop wireless communications to gateways~\cite{LoRa_specification}. As shown in Fig.~\ref{system_model}, the locations of all LoRa nodes are modeled by a PCP, where the LoRa receivers follow the parent point process and the offspring point process (one per parent) are conditionally independent. Specifically, the locations of the LoRa receivers are modeled as a PPP, ${\Phi_G}$, with density ${\lambda_G}$. In each cluster, the locations of the LoRa nodes served by the parent receiver are conditionally independent. The union of all the offspring LoRa nodes constitute a PCP, namely, a \emph{Matern} cluster process.  In this LoRa network, we consider two models for the number of active LoRa nodes distributed in each cluster. One is that the number of active LoRa nodes in each cluster is fixed as $n$. Another is that the number of active LoRa nodes is random and follows a Poisson distribution with average $\bar n$. Different LoRa nodes communicate with the same receiver over the same channel simultaneously by adopting different SFs. Different SFs will lead to different transmission ranges and data rates, for instance, SF 7 can achieve the highest data rate while the transmission distance is limited compared to SF 12.\footnote{How to allocate SFs among different users is out the scope of this paper.}

As in many practical environments, we also assume that there exist non-LoRa nodes transmitting on the same channel, which refer to nodes connecting to other non-LoRa radio modules, as LoRa is working on the unlicensed spectrum. Those nodes are modeled as a PPP, ${\Phi_{\rm co}}$, with density ${\lambda_{\rm co}}$, labeled by diamonds in Fig.~\ref{system_model}, shows  coexistence interference at the LoRa receiver. In a network with LoRa modules only, coexistence interference becomes zero by tuning the density of the non-LoRa nodes as ${\lambda_{\rm co}}=0$.

\subsection{Propagation Model}
As shown in Fig.~\ref{system_model}, for a typical LoRa node, ${U_{x_0}}$, the signal-to-interference-plus-noise ratio (SINR) received at a typical LoRa receiver, $G_{y_0}$, can be expressed as
\begin{equation}\label{SINR_1}
SINR = \frac{{{P_{x_0}}{h_{x_0,y_0}}L\left( {R} \right)}}{{{I_{{\rm{intra}}}} + {I_{{\rm{inter}}}} + {I_{\rm co}} + {\sigma ^2}}},
\end{equation}
where ${I_{{\rm{intra}}}}$, ${I_{{\rm{inter}}}}$, and ${I_{{\rm{co}}}}$ refer to the average powers of intra-interference caused by the non-orthogonality of SFs adopted by LoRa nodes within the same cluster, inter-cluster interference caused by LoRa nodes in the neighboring clusters, and coexistence interference from nodes connecting to non-LoRa radio modules, respectively. These interferences are independent since they are from different sources, and $\sigma^2$ is the power of additive white Gaussian noise (AWGN), and $P_{x_0}$ refers to the transmit power from the typical LoRa node, $U_{x_0}$, and ${h_{x_0,y_0}} \sim \exp \left( 1 \right)$ and $L \left(R\right)$ refer to small-scale fading coefficient and large-scale fading coefficient between the desired LoRa node and its receiver, $G_{y_0}$, respectively. In~\eqref{SINR_1}, $L\left( {R} \right)=\eta {\left\| {{x_0}} \right\|^{ - \alpha }}$, where $\eta$ is a frequency dependent factor, $\left\| {{x_0}} \right\|$ is the distance between the desired LoRa node and its receiver, and $\alpha$ is the path loss exponent.

To get an analytical expression of SINR, we will find ${I_{{\rm{intra}}}}$, ${I_{{\rm{inter}}}}$, and ${I_{{\rm{co}}}}$ subsequently.

\subsubsection{Intra-cluster interference}
At $G_{y_0}$, interference from LoRa nodes within the same cluster except the typical one is
\begin{align}\label{I_intra}
{I_{{\rm{intra}}}} = {\sum _{x \in {N_{y_0}}\backslash {x_0}}}{P_{x}}{h_{x,y_0}}\eta {\left\| x \right\|^{ - \alpha }},
\end{align}
where $P_{x}$ refers to the transmit power of the interference LoRa nodes within the same cluster, and ${h_{x,y_0}}\sim \exp \left( 1 \right)$ is the small-scale fading coefficient, and $\left\| x \right\|$ is the distance between LoRa node and the LoRa receiver in the same cluster, as shown in Fig.~\ref{system_model}. Here, $N_{y_0}$ is the number of the active LoRa nodes that simultaneously transmit to  the same receiver, $G_{y_0}$, within a cluster, which is $n$ for the fixed number of active LoRa nodes model and follows a Poisson distribution with average ${\bar n}$ for the random model.

\subsubsection{Inter-cluster interference}
The interference from LoRa nodes in the adjacent clusters is
\begin{align}
{I_{{\rm{inter}}}} = {\sum _{y \in {\Phi _G}\backslash {y_0}}}\sum_{x \in {N_y}} {{P_{x}}{h_{x,y}}\eta {{\left\| {y + x} \right\|}^{ - \alpha }}},
\label{I_inter}
\end{align}
where $P_{x}$ is the transmit power from the LoRa nodes served by the neighboring LoRa receiver, $G_{y}$, and ${h_{x,y}}\sim \exp \left( 1 \right)$ is the small-scale fading coefficient, and $\left\| {y + x} \right\|$ refers to the distance between the LoRa node, $U_x$, which is served by the neighboring LoRa receivers, $G_{y}$, and the typical LoRa receiver, $G_{y_0}$, as shown in Fig.~\ref{system_model}. $N_y$ is the number of active LoRa nodes served by the LoRa receiver, $G_{y}$, in the same cluster.

\subsubsection{Coexistence interference}
We also consider interference from non-LoRa nodes transmitting over the same frequency, which follows a PPP, ${\Phi_{\rm co}}$, with a density ${\lambda_{\rm co}}$. As such, the interference from these nodes can be expressed as
\begin{align}\label{I_co}
{I_{\rm co}} = {\sum _{z \in {\Phi _{{\rm{co}}}}}}{P_{z}}{h_{z,y_0}}\eta {\left\| z \right\|^{ - \alpha }},
\end{align}
where ${P_{z}}$ is the transmit power from the non-LoRa node, $U_z$, served by non-LoRa radio modules, the distance between the LoRa node, $U_z$, and the typical LoRa receiver, $G_{y_0}$, is $\left\| z \right\|$, and ${h_{z,{y_0}}} \sim \exp\left( 1 \right)$ is small-scale fading coefficient, as shown in Fig.~\ref{system_model}.

\section{Interference Analysis}
In this section, we analyze the intra interference, inter interference and coexistence interference for the models with a fixed and a random number of LoRa nodes in each cluster, where we consider both the ordered case and unordered case. The \emph{unordered case} refers to that the typical LoRa node in the typical cluster is chosen randomly from the set of active LoRa nodes in one cluster. The \emph{ordered case} refers to that the typical LoRa node  is located at the $k$-th closest distance to the LoRa receiver among the active nodes in one cluster. Therefore, there are four scenarios. In this section, we are going to investigate the interference for the four scenarios, respectively.

\subsection{Distance Distributions}
\subsubsection{Unordered distance PDF for intra-cluster} For a \emph{Matern} cluster process, the probability density function (PDF) of the distance, $r$, between a typical receiver centered at a cluster and a connected LoRa node is \cite{Martin_TIF_2009}
\begin{align}\label{PDF_Rayleigh}
{f_{\left\| x \right\|}}\left( x \right) = \left\{ {\begin{array}{*{20}{l}}
{\frac{1}{{\pi {a^2}}},{\mkern 1mu} \left\| x \right\| \le a}.\\
{0,{\mkern 1mu} {\rm{otherwise}}},
\end{array}} \right.
\end{align}
where $a$ is the cluster radius. When converting~\eqref{PDF_Rayleigh} to polar coordinates, we have
\begin{align}\label{PDF_polar}
{f_R}\left( r \right) = \left\{ {\begin{array}{*{20}{l}}
{\frac{{2r}}{{{a^2}}},{\mkern 1mu} r \le a}.\\
{0,{\mkern 1mu} {\rm{otherwise}}}.
\end{array}} \right.
\end{align}

\subsubsection{Ordered distance PDF for intra-cluster}
In the ordered case, the PDF for intra-cluster can be given by the following lemma.
\begin{lemma}\label{Order PDF intra}
If $n_\iota$ active LoRa nodes are uniformly distributed within a cluster with radius, $a$, then the PDF of the distance of the $k$-th closest LoRa node connecting to the typical LoRa receiver is given by
\begin{align}\label{Order PDF intra 1}
{f_{{{\tilde r}_k}}}\left( r \right) = \left\{ \begin{array}{l}
\frac{{2{n_\iota }!\sum\nolimits_{p = 0}^{{n_\iota } - k} {{{\left( { - 1} \right)}^p}{a^{ - 2\left( {p + k} \right)}}{r^{2\left( {p + k} \right) - 1}}} }}{{\left( {{n_\iota } - k} \right)!\left( {k - 1} \right)!}}, r \le a.\\
0,\;\;{\rm{otherwise}},
\end{array} \right.
\end{align}
where ${n_\iota } \in \left\{ {n,\bar n} \right\}$.

\begin{proof}
With the aid of order statistics \cite{order}, the ordered distance PDF is given by
\begin{align}\label{Order PDF intra_2}
{f_{{{\tilde r}_k}}}\left( r \right) = \frac{{{n_\iota }!{{\left( {{F_R}\left( r \right)} \right)}^{k - 1}}{{\left( {1 - {F_R}\left( r \right)} \right)}^{{n_\iota } - k}}{f_R}\left( r \right)}}{{\left( {{n_\iota } - k} \right)!\left( {k - 1} \right)!}},
\end{align}
where
\begin{align}\label{Order CDF intra_2}
{F_R}\left( r \right) = \frac{{{r^2}}}{{{a^2}}}
\end{align}
is the CDF of the unordered intra-cluster distance. Substituting ${F_R}\left( r \right)$ and \eqref{PDF_polar} into \eqref{Order PDF intra_2} and applying binomial series expansion, we can obtain \eqref{Order PDF intra 1}. The proof is completed.
\end{proof}
\end{lemma}

\subsubsection{PDF of the ordered distance for intra-cluster interference LoRa nodes}
Since we are interested in the performance of the $k$-th closest LoRa node, its corresponding intra-cluster interference nodes are dependent on the distance rank $k$. To better incorporate our analytical procedure, besides the $k$-th LoRa node, we further divide the rest $\left( {{n_\iota } - 1} \right)$ nodes into two sets. Those closer than the $k$-th one,  ${\mathcal{K}_{near}} = \left\{ {1, \cdots ,k - 1} \right\}$,  and those farther away than the $k$-th one, ${\mathcal{K}_{far}} = \left\{ {k + 1, \cdots ,{n_\iota }} \right\}$.

\begin{lemma}\label{Order PDF intra interfering}
The PDF of the ordered LoRa nodes that interfere the $k$-th closest node  in the same cluster is given by
\begin{align}\label{Order PDF intra interfering 1}
{f_{\tilde R}}\left( {\left. r \right|{{\tilde r}_k}} \right) = \left\{ \begin{array}{l}
\frac{{2r}}{{{{\tilde r}_k}^2}},r \le {{\tilde r}_k}.\\
\frac{{2r}}{{{a^2} - {{\tilde r}_k}^2}},{{\tilde r}_k} < r \le a,
\end{array} \right.
\end{align}
where ${\tilde r}_k$ denotes the distance from the $k$-th closest LoRa node to the serving receiver.
\begin{proof}
Following the same procedure for obtaining Lemma 4 in~\cite{Afshang_TWC:2016}, with the aid of ordered statistics \eqref{Order PDF intra 1} and applying the symmetry property, we can obtain the distance PDF of the ordered intra-cluster for ${\mathcal{K}_{near}}$ and ${\mathcal{K}_{far}}$ as ${f_{{r_{near}}}}\left( {\left. {{r_{near}}} \right|{{\tilde r}_k}} \right) = \frac{{{f_R}\left( {{r_{near}}} \right)}}{{{F_R}\left( {{{\tilde r}_k}} \right)}}$ and ${f_{{r_{far}}}}\left( {\left. {{r_{far}}} \right|{{\tilde r}_k}} \right) = \frac{{{f_R}\left( {{r_{far}}} \right)}}{{1 - {F_R}\left( {{{\tilde r}_k}} \right)}}$, respectively. Then with the aid of \eqref{Order PDF intra_2} and \eqref{Order CDF intra_2}, we can obtain the desired result in~\eqref{Order PDF intra interfering 1}. The proof is completed.
\end{proof}
\end{lemma}

\subsection{Laplace Transforms for Interferences of Unordered LoRa Nodes}
In the following, we turn our attention to obtaining the Laplace transforms of the interference parts for the unordered cases with a fixed and a random number of active LoRa nodes in each cluster, respectively.

\subsubsection{Intra-cluster interference}
We first address the intra-cluster interference.
\begin{lemma}\label{Intra Laplace unordered fixed n}
For the unordered case with a fixed number of active nodes $n$ in each cluster, the Laplace transform of intra-cluster interference can be expressed as
\begin{align}\label{Intra Laplace unordered fixed n 1}
{{\cal L}_{I_{{\rm{intra}}}^n}}\left( s \right) = {\left[ {\frac{{{a^\alpha }\delta }}{{s{P_x}\eta \left( {\delta  + 1} \right)}}{}_2{F_1}\left( {1,\delta  + 1;\delta  + 2; - \frac{{{a^\alpha }}}{{s{P_x}\eta }}} \right)} \right]^{n - 1}},
\end{align}
and for the unordered case with with a random number of active LoRa nodes of average ${\bar n}$ in the cluster, the Laplace transform of intra-cluster interference can be expressed as
\begin{align}\label{Intra Laplace 1}
{{\cal L}_{I_{{\rm{intra}}}^{\bar n}}}\left( s \right) = \exp \left( { - \left( {\bar n - 1} \right){{{}_2{F_1}}}\left( {1,\delta ;\delta  + 1; - \frac{{{a^\alpha }}}{{s{P_x}\eta }}} \right)} \right),
\end{align}
 where ${}_2{F_1}\left(\right)$ is the is the Gauss hypergeometric function.

\begin{proof}
See Appendix A.
\end{proof}
\end{lemma}

To obtain more insightful and simple expressions, we use Gauss-Chebyshev approximation, which is regarded as a tight approximation and has been widely used~\cite{Hildebrand1987introduction,Ding:SPL:2014,yuanwei_JSAC_2015,LV:TVT:2017,LIU:TWC:2017}, to approximate~\eqref{Intra Laplace unordered fixed n 1} and~\eqref{Intra Laplace 1}. It is shown by the following corollary.
\begin{corollary}\label{Intra Laplace fixed n app}
For the  unordered case with a fixed number of active LoRa nodes in the cluster, the Laplace transform of the intra-cluster interference  can be approximated as
\begin{align}\label{Intra Laplace fixed n app 1}
\mathcal{L}_{I_{{\rm{intra}}}^n}^{ap}\left( s \right) \approx {\left[ {{\omega _T}\sum\limits_{t = 1}^T {\frac{{{\mu _t}c_t^{\alpha  + 1}}}{{c_t^\alpha  + {a^{ - \alpha }}s{P_x}\eta }}} } \right]^{n - 1}},
\end{align}
and for the unordered case with a random number of active LoRa nodes with average $\bar n$ in the cluster, the Laplace transform of the intra-cluster interference can be approximated as
\begin{align}\label{Intra Laplace app 1}
{{\cal L}_{I_{{\rm{intra}}}^{\bar n,ap}}}\left( s \right) \approx \exp \left( { - \left( {\bar n - 1} \right){\omega _T}\sum\limits_{t = 1}^T {\frac{{{\mu _t}{c_t}}}{{\frac{{c_t^\alpha {a^\alpha }}}{{s{P_{x}}\eta }} + 1}}} } \right),
\end{align}
where ${\omega _T} = \frac{\pi }{T},{\psi _t} = \cos \left( {\frac{{2t - 1}}{{2T}}\pi } \right),{c_t} = \frac{{\left( {{\psi _t} + 1} \right)}}{2}$, and ${\mu _t} = \sqrt {1 - \psi _t^2} $.

\end{corollary}

\subsubsection{Inter-cluster interference}
Here, we provide the Laplace transforms of the inter-cluster interference for a fixed and a random number of active LoRa nodes in the cluster, respectively.

\begin{lemma}\label{Inter Laplace fixed lower bound}
The Laplace transform of the inter-cluster interference with a fixed number of active nodes $n$ in the cluster is upper bounded by
\begin{align}\label{Inter Laplace fixed lower bound_1}
&L_{I_{{\rm{inter}}}^n}^{up}\left( s \right) = \\ \nonumber &exp \left( { - \pi {\lambda _G}{{\left( {s{P_x}\eta } \right)}^\delta }\delta \sum\limits_{p = 1}^n {
n\choose
p
} B\left( {p - \delta ,n - p + \delta } \right)} \right),
\end{align}
where $B\left( { \cdot , \cdot } \right)$ is the Beta function. Here, $B\left( {x,y} \right) = \int_0^\infty  {\frac{{{t^{x - 1}}}}{{{{\left( {1 + t} \right)}^{x + y}}}}} dt$ for $x,y > 0$.

The Laplace transform of the inter-cluster interference with a random number of active nodes of average $\bar n$ in a cluster is lower bounded by
\begin{align}\label{Inter Laplace final_appro_align}
\mathcal{L}_{I_{{\rm{inter}}}^{\bar n}}^{low}\left( s \right) = \exp \left( { - {\pi ^2}{\lambda _G}\bar n{{\left( {s{P_x}\eta } \right)}^\delta }\frac{\delta }{{\sin \left( {\pi \delta } \right)}}} \right).
\end{align}
\begin{proof}
For the proof of~\eqref{Inter Laplace fixed lower bound_1}, see Appendix~B. For the proof of~\eqref{Inter Laplace final_appro_align}, see Appendix~C.
\end{proof}
\end{lemma}

\subsubsection{Coexistence interference} For coexistence interference, we can obtain the following lemma.
\begin{lemma}\label{Coexistence Laplace}
The Laplace transform of coexistence interference, which comes from the nodes connecting to non-LoRa radio modules, can be expressed as
\begin{align}\label{Coexistence Laplace 1}
{{\cal L}_{{I_{{\rm{co}}}}}}\left( s \right) = \exp \left( { - \pi{\lambda _{{\rm{co}}}} \Gamma \left( {1 + \delta } \right)\Gamma \left( {1 - \delta } \right){{\left( {s{P_{z}}\eta } \right)}^\delta }} \right),
\end{align}
where $\delta  = \frac{2}{\alpha }$.
\begin{proof}
\begin{align}\label{Coexistence proof 1}
{{\cal L}_{{I_{{\rm{co}}}}}}\left( s \right) &= {\mathbb{E}_{{I_{{\rm{co}}}}}}\left\{ {\exp \left( { - s\sum\limits_{z \in {\Phi _{{\rm{co}}}}} {{P_{z}}{h_{z,y_0}}\eta {{{\left\| z \right\|}}^{ - \alpha }}} } \right)} \right\}\nonumber\\
&\mathop {\rm{ = }}\limits^{\left( a \right)}  {\mathbb{E}_{{\Phi _{{\rm{co}}}}}}\left\{ {\prod\limits_{z \in {\Phi _{{\rm{co}}}}} {{\mathbb{E}_{{h_{z,y_0}}}}\left[ {\exp \left( { - s{P_{z}}{h_{z,y_0}}\eta {{{\left\| z \right\|}}^{ - \alpha }}} \right)} \right]} } \right\}\nonumber\\
&\mathop  = \limits^{\left( b \right)} \exp \left( { - 2\pi{\lambda _{{\rm{co}}}} \int_0^\infty  {\left( {1 - \frac{1}{{1 + s{P_{z}}\eta {r^{ - \alpha }}}}} \right)rdr} } \right),
\end{align}
where $(a)$ is obtained by applying the generating function, and $(b)$ is obtained by the fact that ${{g_{z,y_0}}}$ follows a Rayleigh fading distribution. By applying \cite[ Eq. (3.241) .4]{gradshteyn}, we can obtain \eqref{Coexistence Laplace 1}.  The proof is completed.
\end{proof}
\end{lemma}

\subsection{Laplace Transforms for Interferences of Ordered LoRa Users}

\begin{lemma}\label{Intra Laplace ordered fixed number}
For the ordered case with a fixed number of active nodes $n$ in the cluster, the Laplace transform of intra-cluster interference  can be expressed as
\begin{align}\label{Intra Laplace ordered fixed number 1}
&{{\cal L}_{I_{{\rm{intra}}}^{n,k}}}\left( s \right) = {\left[ {\frac{{{{\tilde r}_k}^\alpha \delta }}{{s{P_x}\eta \left( {\delta  + 1} \right)}}{Z_1}\left( {{{\tilde r}_k}} \right)} \right]^{k - 1}}  \\ \nonumber &{\left[ {\frac{\delta }{{\left( {{a^2} - {{\tilde r}_k}^2} \right)s{P_x}\eta \left( {\delta  + 1} \right)}}\left( {{a^{\alpha  + 2}}{Z_1}\left( a \right) - {{\tilde r}_k}^{\alpha  + 2}{Z_1}\left( {{{\tilde r}_k}} \right)} \right)} \right]^{n - k}},
\end{align}
where
\begin{align}
{Z_1}\left( {{{\tilde r}_k}} \right) = {}_2{F_1}\left( {1,\delta  + 1;\delta  + 2; - {{\left( {s{P_x}\eta } \right)}^{ - 1}}{{\tilde r}_k}^\alpha } \right),
\end{align} and
\begin{align}
{Z_1}\left( a \right) = {}_2{F_1}\left( {1,\delta  + 1;\delta  + 2; - {{\left( {s{P_x}\eta } \right)}^{ - 1}}{a^\alpha }} \right).
\end{align}

For the ordered case with a random number of active LoRa nodes in the cluster, with the number of active LoRa users, $\mathcal{N}$, no fewer than the average number, $\bar n$, the corresponding Laplace transform of intra-cluster interference can be expressed as
\begin{align}\label{Intra Laplace ordered worst 1}
{{\cal L}_{I_{{\rm{intra}}}^{\bar n,\mathcal{N}}}}\left( s \right) =& \exp \left[ { - {{\left( {\bar n - 1} \right)}_2}{F_1}\left( {1,\delta ;\delta  + 1; - {{\left( {s{P_x}\eta } \right)}^{ - 1}}{{\tilde r}_\mathcal{N}}^\alpha } \right)} \right],
\end{align}
where ${\tilde r}_\mathcal{N}$ is the distance from the LoRa node to the serving receiver.
\begin{proof}
For the proof, see Appendix~D.
\end{proof}
\end{lemma}

From Lemma~\ref{Intra Laplace ordered fixed number} and using Gauss-Chebyshev approximation, we can obtain the following corollary.

\begin{corollary}\label{Intra fixed ordered GC}
For the ordered case with a fixed number of active LoRa nodes $n$  in the cluster , the Laplace transform of intra-cluster interference can be approximated as
\begin{align}\label{Intra fixed ordered GC_1}
&\mathcal{L}_{I_{{\rm{intra}}}^{n,k}}^{ap}\left( s \right)  \approx  {\left[ {{\omega _T}\sum\limits_{t = 1}^T {\frac{{{\mu _t}c_t^{\alpha  + 1}}}{{c_t^\alpha  + {{\tilde r}_k}^{ - \alpha }s{P_x}\eta }}} } \right]^{k - 1}} \times\\ \nonumber
&{\left[ {\frac{{{\omega _T}}}{{{a^2} - {{\tilde r}_k}^2}}\left( {{a^2}\sum\limits_{t = 1}^T {\frac{{{\mu _t}c_t^{\alpha  + 1}}}{{c_t^\alpha  + {a^{ - \alpha }}s{P_x}\eta }}}  - {{\tilde r}_k}^2\sum\limits_{t = 1}^T {\frac{{{\mu _t}c_t^{\alpha  + 1}}}{{c_t^\alpha  + {{\tilde r}_k}^{ - \alpha }s{P_x}\eta }}} } \right)} \right]^{n - k}},
\end{align}
and for the ordered case with a random number of active LoRa nodes with average $\bar n$ in the cluster, the Laplace transform of intra-cluster interference  can be approximated as
\begin{align}\label{Intra mean ordered GC_1}
&\mathcal{L}_{I_{{\rm{intra}}}^{\bar n,\mathcal{N}}}^{ap}\left( s \right) \approx \exp \left( { - \left( {\bar n - 1} \right){\omega _T}\sum\limits_{t = 1}^T {\frac{{{\mu _t}{c_t}}}{{\frac{{c_t^\alpha {{\tilde r}_\mathcal{N}}^\alpha }}{{s{P_x}\eta }} + 1}}} } \right).
\end{align}

\end{corollary}

Note that the Laplace transforms of inter-cluster interference and coexistence interference for the ordered case are the same as those in the unordered case, which are given in the last subsection.

\section{Coverage Probability Analysis}
In this section, we derive the coverage probability for a typical LoRa node based on the interference analysis in the previous section. The coverage probability is defined as the probability that a typical user can be successfully decoded  at the receiver, that is SINR of the typical user received  at the receiver is higher than a threshold, $\gamma_{th}$.  It can be expressed as
\begin{align}\label{SINR_cov}
{P_{{\rm{cov}}}}\left( {{\gamma _{th}}} \right) = {\mathbb{E}_R}\left[ {\left. {\Pr \left\{ {\mathrm{SINR}\left( R \right) \ge {\gamma _{th}}} \right\}} \right|R} \right],
\end{align}
where $R$ is the distance between the desired LoRa node and the serving receiver in the same cluster.

By applying \eqref{SINR_1} and~\eqref{PDF_Rayleigh} into \eqref{SINR_cov}, the transmission coverage probability of a typical LoRa node can be expressed as
\begin{align}\label{SINR}
&{P_{{\rm{cov}}}}\left( {{\gamma _{th}}} \right) =
\int {\Pr } \left\{ {\mathrm{SINR}\left( R \right) \ge {\gamma _{th}}} \right\}{f_R}\left( r \right)dr.
\end{align}
From~\eqref{SINR_cov}, we can obtain
\begin{align}\label{coverage temp 1}
&\Pr \left\{ {\mathrm{SINR}\left( R \right) \ge {\gamma _{th}}} \right\} = \Pr \left\{ {\frac{{{P_{{x_0}}}{h_{{x_0},{y_0}}}L\left( {R} \right)}}{{{I_{{\rm{intra}}}} + {I_{{\rm{inter}}}} + {I_{{\rm{co}}}} + {\sigma ^2}}} \ge {\gamma _{th}}} \right\}\nonumber\\
&  = \Pr \left\{ {{h_{{x_0},{y_0}}} \ge \left( {{I_{{\rm{intra}}}} + {I_{{\rm{inter}}}} + {I_{co}} + {\sigma ^2}} \right)\frac{{\eta {R^\alpha }{\gamma _{th}}}}{{{P_{{x_0}}}}}} \right\}\nonumber\\
&= {e^{ - \rho{\sigma^2}}}{\mathbb{E}_{{I_{{\rm{intra}}}} + {I_{{\rm{inter}}}} + {I_{\rm co}}}}\left\{ {e^{ - \rho\left( {{I_{{\mathop{\rm intra}} }} + {I_{{\mathop{\rm inter}} }} + {I_{\rm co}}} \right) }} \right\}\nonumber\\
&= {e^{ - \rho{\sigma^2}}}{{\cal L}_{{I_{{\mathop{\rm intra}} }}}}\left( \rho \right){{\cal L}_{{I_{{\mathop{\rm inter}} }}}}\left( \rho \right){{\cal L}_{{I_{\rm co}}}}\left( \rho \right),
\end{align}
where $\rho={\frac{{R^\alpha {\gamma _{th}}}}{{{P_{x_0}}\eta }}}$, ${{\cal L}_{{I_{{\mathop{\rm intra}}}}}}\left( \rho  \right) = \mathbb{E}\left\{ {{e^{- \rho {I_{{\mathop{\rm intra}}}}}}} \right\}$, ${{\cal L}_{{I_{{\mathop{\rm inter}}}}}}\left( \rho  \right) = \mathbb{E}\left\{ {{e^{ -\rho {I_{{\mathop{\rm inter}}}}}}} \right\}$, and ${{\cal L}_{{I_{{\mathop{\rm co}}}}}}\left( \rho  \right) = \mathbb{E}\left\{ {{e^{ -\rho {I_{{\mathop{\rm co}}}}}}} \right\}$ are the Laplace transforms of the power density distributions of ${{I_{{\mathop{\rm intra}}}}}$, ${{I_{{\mathop{\rm inter}}}}}$ and ${{I_{{\mathop{\rm co}}}}}$, respectively.

\subsection{Coverage Probability of Unordered LoRa Users}
Based on the derived results in Section III, we can obtain the coverage probability of a typical LoRa node for the unordered case in this subsection.
\begin{theorem}\label{coverage provability exact}
If  the LoRa nodes follow a Matern cluster process centered around each receiver, for the unordered case with a fixed number of LoRa nodes in the cluster, the upper bound of the coverage probability of a typical LoRa node can be expressed as
\begin{align}\label{coverage provability exact 1}
P_{{\rm{cov,up}}}^n\left( {{\gamma _{th}}} \right) = \frac{2}{{{a^2}}}\int_0^a {{e^{ - \rho {\sigma ^2}}}{\mathcal{L}_{I_{{\rm{intra}}}^n}}\left( \rho  \right)} \mathcal{L}_{I_{{\rm{inter}}}^{n}}^{up}\left( \rho  \right){\mathcal{L}_{{I_{{\rm{co}}}}}}\left( \rho  \right)rdr,
\end{align}
where ${{\mathcal{L}_{I_{{\rm{intra}}}^n}}\left( \rho  \right)}$, $\mathcal{L}_{I_{{\rm{inter}}}^{n}}^{up}\left( \rho  \right)$, and ${{\cal L}_{{I_{{\rm{co}}}}}}\left( {{\rho }} \right)$ are given by \eqref{Intra Laplace unordered fixed n 1}, \eqref{Inter Laplace fixed lower bound_1}, and \eqref{Coexistence Laplace 1}.

For the unordered case with a random number of LoRa nodes in the cluster with average $\bar n$, the lower bound of the coverage probability of a typical LoRa node can be expressed as
\begin{align}\label{coverage provability mean exact 1}
P_{{\rm{cov,low}}}^{\bar n}\left( {{\gamma _{th}}} \right) = \frac{2}{{{a^2}}}\int_0^a {{e^{ - \rho {\sigma ^2}}}{{\cal L}_{I_{{\rm{intra}}}^{\bar n}}}\left( \rho  \right)} {\cal L}_{I_{{\rm{inter}}}^{\bar n}}^{low}\left( \rho  \right){{\cal L}_{{I_{{\rm{co}}}}}}\left( \rho  \right)rdr,
\end{align}
where ${{{\cal L}_{I_{{\rm{intra}}}^{\bar n}}}\left( \rho  \right)}$, ${\cal L}_{I_{{\rm{inter}}}^{\bar n}}^{low}\left( \rho  \right)$, and ${{\cal L}_{{I_{{\rm{co}}}}}}\left( {{\rho }} \right)$ are given by \eqref{Intra Laplace 1}, \eqref{Inter Laplace final_appro_align}, and \eqref{Coexistence Laplace 1}.

\begin{proof}
Substituting \eqref{PDF_polar} and \eqref{coverage temp 1} into \eqref{SINR}, and basing the derived results of  Lemmas \ref{Intra Laplace unordered fixed n}, \ref{Inter Laplace fixed lower bound} and \ref{Coexistence Laplace}, we can obtain the desired result in \eqref{coverage provability exact 1}.  The proof procedure of obtaining~\eqref{coverage provability mean exact 1} is similar to the above for~\eqref{coverage provability exact 1}, and is hence skipped here.
\end{proof}
\end{theorem}

Note that it is hard to see insights directly from~\eqref{coverage provability exact 1} and \eqref{coverage provability mean exact 1}. To obtain more insightful expressions, we derive the following corollary.

\begin{corollary}\label{coverage provability fixed closed-form}
If the LoRa nodes follow a Matern cluster process centered around each receiver, for the unordered case with a fixed number of LoRa nodes in the cluster,  the coverage probability of a typical LoRa node can be approximated  as
\begin{align}\label{coverage provability fixed closed-form 1}
&P_{{\rm{cov,up}}}^{n,ap}\left( {{\gamma _{th}}} \right)\approx \nonumber \\
& {\omega _M}\sum\limits_{m = 1}^M {{\vartheta _m}{l_m}\exp \left( { - {\rho _m}{\sigma ^2}} \right){{\left[ {{\omega _T}\sum\limits_{t = 1}^T {\frac{{{\mu _t}c_t^{\alpha  + 1}}}{{c_t^\alpha  + {\gamma _{th}}{{\tilde P}_x}^{ - 1}}}} } \right]}^{n - 1}}} \nonumber\\
& \exp \left[ { - {a^2}l_m^2\pi {\lambda _G}\delta \sum\limits_{p = 1}^n {
n\choose
p} B\left( {p - \delta ,n - p + \delta } \right){{\left( {{\gamma _{th}}{{\tilde P}_x}^{ - 1}} \right)}^\delta }} \right] \nonumber\\
&\exp \left[ { - \frac{{{a^2}l_m^2{\pi ^2}\delta }}{{\sin \left( {\pi \delta } \right)}}{\lambda _{{\rm{co}}}}{{\left( {{\gamma _{th}}{{\tilde P}_z}^{ - 1}} \right)}^\delta }} \right],
\end{align}
where ${{\tilde P}_{x}} = \frac{{{P_{x_0}}}}{{{P_{x}}}},{{\tilde P}_{z}} = \frac{{{P_{x_0}}}}{{{P_{z}}}}$, ${\omega _M} = \frac{\pi }{M}$, ${\nu _m} = \cos \left( {\frac{{2m - 1}}{{2M}}\pi } \right)$, ${l_m} = \frac{{\left( {{\nu _m} + 1} \right)}}{2}$, ${\rho _m} = \frac{{l_m^\alpha {a^\alpha }{\gamma _{th}}}}{{{P_{{x_0}}}\eta }}$, and ${\vartheta _m} = \sqrt {1 - \nu _m^2}$, ${\omega _T} = \frac{\pi }{T},{\psi _t} = \cos \left( {\frac{{2t - 1}}{{2T}}\pi } \right),{c_t} = \frac{{\left( {{\psi _t} + 1} \right)}}{2}$, and ${\mu _t} = \sqrt {1 - \psi _t^2} $.

For the unordered case with a random number of active LoRa nodes in the cluster with average $\bar n$, the coverage probability of a typical LoRa node  can be approximated as
\begin{align}\label{coverage provability closed-form 1}
&P_{{\rm{cov,low}}}^{\bar n,ap}\left( {{\gamma _{th}}} \right) \approx {\omega _M}\sum\limits_{m = 1}^M {{\vartheta _m}{l_m}\exp \left\{ { - {\rho _m}{\sigma ^2} - \left( {\bar n - 1} \right){\omega _T}} \right.} \nonumber\\
&\left. {\sum\limits_{t = 1}^T {\frac{{{\mu _t}{c_t}}}{{\frac{{c_t^\alpha {{\tilde P}_{x}}}}{{l_m^\alpha {\gamma _{th}}}} + 1}}}  - \frac{{{a^2}l_m^2{\pi ^2}\delta }}{{\sin \left( {\pi \delta } \right)}}\left( {{\lambda _G}\bar n{{\left( {\frac{{{\gamma _{th}}}}{{{{\tilde P}_{x}}}}} \right)}^\delta } + {\lambda _{{\rm{co}}}}{{\left( {\frac{{{\gamma _{th}}}}{{{{\tilde P}_{z}}}}} \right)}^\delta }} \right)} \right\}.
\end{align}

\begin{proof}
For simplicity, by using the approximated expressions of the Laplace transform rather than the exact expressions, we can obtain
\begin{align}\label{coverage fixed provability appro 2}
&P_{{\rm{cov,up}}}^{n,ap}\left( {{\gamma _{th}}} \right) \approx \frac{2}{{{a^2}}}\int_0^a {{e^{ - \rho {\sigma ^2}}}{\cal L}_{I_{{\rm{intra}}}^n}^{ap}\left( \rho  \right)} \mathcal{L}_{I_{{\rm{inter}}}^{n}}^{up}\left( \rho  \right){\mathcal{L}_{{I_{{\rm{co}}}}}}\left( \rho  \right)rdr.
\end{align}
Then based on the derived results of  Corollary \ref{Intra Laplace fixed n app},  Lemmas \ref{Inter Laplace fixed lower bound} and \ref{Coexistence Laplace}, with the aid of Gauss-Chebyshev approximation, we can obtain the desired results in closed-form as \eqref{coverage provability fixed closed-form 1}.  The proof procedure to obtain~\eqref{coverage provability closed-form 1} is similar to the above for~\eqref{coverage provability fixed closed-form 1}, and is hence skipped here.
\end{proof}
\end{corollary}

\begin{remark}\label{coverage provability remark 1}
The coverage probability is a monotonic decreasing function of ${n_\iota } \in \left\{ {n,\bar n} \right\}$, ${{\lambda _G}}$, and ${{\lambda_{\rm co}}}$. This indicates that increasing the average number of LoRa nodes that simultaneously transmit to the same LoRa receiver degrades the coverage probability. It is further observed that one  either increases the density of active LoRa nodes, or the density of active non-LoRa nodes degrades the coverage probability, as more intra-interference and inter-interference are introduced.
\end{remark}

\begin{remark}\label{coverage provability remark 2}
The coverage probability is a monotonic decreasing function of the cluster radius, $a$. Hence, if the LoRa nodes are densely deployed around each LoRa receiver, the coverage probability of LoRa nodes will be enhanced.
\end{remark}

If the considered networks are \emph{intra-interference limited}, that is, $I_{\rm inter}=I_{\rm co}=0$ and ${\sigma ^2} = 0$, then from~\eqref{coverage provability fixed closed-form 1} and~\eqref{coverage provability closed-form 1}, we can obtain the following propagation.

\begin{proposition}\label{coverage provability intra only fixed}
For the unordered intra-interference limited case, the coverage probability of the case with a fixed number of active nodes $n$ in the cluster can be expressed   as
\begin{align}\label{coverage provability intra only fixed 1}
P_{{\rm{intra,up}}}^{n,ap}\left( {{\gamma _{th}}} \right) \approx {\omega _M}\sum\limits_{m = 1}^M {{\vartheta _m}{l_m}{{\left[ {{\omega _T}\sum\limits_{t = 1}^T {\frac{{{\mu _t}c_t^{\alpha  + 1}}}{{c_t^\alpha  + {\gamma _{th}}{{\tilde P}_x}^{ - 1}}}} } \right]}^{n - 1}}} ,
\end{align}
and the coverage probability of the case with a random number of active LoRa nodes  in the cluster with average $\bar n$ can be expressed as
\begin{align}\label{coverage provability intra only 1}
&P_{{\rm{intra,low}}}^{\bar n,ap}\left( {{\gamma _{th}}} \right)\approx \nonumber\\
& {\omega _M}\sum\limits_{m = 1}^M {{\vartheta _m}{l_m}\exp \left( { - \left( {\bar n - 1} \right)\sum\limits_{t = 1}^T {\frac{{{\omega _T}{\mu _t}{c_t}}}{{\frac{{c_t^\alpha {{\tilde P}_x}}}{{l_m^\alpha {\gamma _{th}}}} + 1}}} } \right)}.
\end{align}
\end{proposition}
\begin{remark}\label{coverage provability intra only 2}
When the networks are intra-interference limited, from ~\eqref{coverage provability fixed closed-form 1} and~\eqref{coverage provability intra only 1}, the coverage probability is independent of the cluster radius, $a$, which is intuitive.
\end{remark}

\subsection{Coverage Probability of Ordered LoRa Users}
Based on the derived results in Section III, we can obtain the coverage probability of the $k$-th closest LoRa node in the ordered case in this subsection.

The transmission coverage probability of the $k$-th closest LoRa node can be expressed as
\begin{align}\label{coverage probability k nearest_1}
P_{{\mathop{\rm cov}} }^k\left( {{\gamma _{th}}} \right) = \int_{{R^ + }} {\Pr \left\{ {{\rm{SINR}}\left( {{{\tilde r}_k}} \right) \ge {\gamma _{th}}} \right\}{f_{{{\tilde r}_k}}}\left( {{{\tilde r}_k}} \right)d{{\tilde r}_k}},
\end{align}

\begin{theorem}\label{coverage provability k nearest ordered}
If the LoRa nodes follow a Matern cluster process centered around each receiver,  the coverage probability of the $k$-th closest LoRa node located in a cluster of $n$ active LoRa nodes can be expressed as
\begin{align}\label{coverage provability k nearest ordered 1}
P_{{\rm{cov,up}}}^{n,k}\left( {{\gamma _{th}}} \right) &= \int_0^a {{e^{ - {\rho _k}{\sigma ^2}}}{L_{I_{{\rm{intra}}}^{n,k}}}\left( {{\rho _k}} \right)} L_{I_{{\rm{inter}}}^n}^{up}\left( {{\rho _k}} \right){L_{{I_{{\rm{co}}}}}}\left( {{\rho _k}} \right) \nonumber\\& \times \frac{{2n!{{\tilde r}_k}^{2k - 1}{{\left( {1 - {{\tilde r}_k}^2{a^{ - 2}}} \right)}^{n - k}}}}{{{a^{2k}}\left( {n - k} \right)!\left( {k - 1} \right)!}}d{{\tilde r}_k},
\end{align}
where ${\rho _k} = \frac{{{{\tilde r}_k}^\alpha {\gamma _{th}}}}{{{P_{{x_0}}}\eta }}$, ${{{\cal L}_{I_{{\rm{intra}}}^{n,k}}}\left( {{\rho _k}} \right)}$, $\mathcal{L}_{I_{{\rm{inter}}}^{n}}^{up}\left( \rho  \right)$, and ${{\cal L}_{{I_{{\rm{co}}}}}}\left( {{\rho }} \right)$ are given by \eqref{Intra Laplace ordered fixed number 1}, \eqref{Inter Laplace fixed lower bound_1}, and \eqref{Coexistence Laplace 1}.

The coverage probability of the $k$-th closest LoRa node located in a cluster with a random number of active nodes with average $\bar n$ can be expressed as
\begin{align}\label{coverage provability k nearest ordered mean 1}
P_{{\rm{cov,low}}}^{\bar n,k}\left( {{\gamma _{th}}} \right) &= \int_0^a {{e^{ - {\rho _k}{\sigma ^2}}}{L_{I_{{\rm{intra}}}^{\bar n,k}}}\left( {{\rho _k}} \right)} L_{I_{{\rm{inter}}}^{\bar n}}^{low}\left( {{\rho _k}} \right){L_{{I_{{\rm{co}}}}}}\left( {{\rho _k}} \right)\nonumber\\ &\times\frac{{2n!{{\tilde r}_k}^{2k - 1}{{\left( {1 - {{\tilde r}_k}^2{a^{ - 2}}} \right)}^{n - k}}}}{{{a^{2k}}\left( {n - k} \right)!\left( {k - 1} \right)!}}d{{\tilde r}_k},
\end{align}
where ${\rho _k} = \frac{{{{\tilde r}_k}^\alpha {\gamma _{th}}}}{{{P_{{x_0}}}\eta }}$, ${{{\cal L}_{I_{{\rm{intra}}}^{\bar n,k}}}\left( {{\rho _k}} \right)}$, $\mathcal{L}_{I_{{\rm{inter}}}^{\bar n}}^{low}\left( \rho_k  \right)$, and ${{\cal L}_{{I_{{\rm{co}}}}}}\left( {{\rho_k }} \right)$ are given by \eqref{Intra Laplace ordered worst 1}, \eqref{Inter Laplace final_appro_align}, and \eqref{Coexistence Laplace 1}.
\begin{proof}
By applying~\eqref{Order PDF intra 1} and \eqref{SINR_1}  into \eqref{coverage probability k nearest_1}, we obtain the desired results. The proof is completed.
\end{proof}
\end{theorem}

\begin{corollary}\label{coverage provability fixed n GC closed-form}
If the LoRa nodes follow a Matern cluster process  centered around each receiver, the coverage probability of the $k$-th closest LoRa node, which is located in a cluster with a fixed number of active nodes $n$, can be approximated as
\eqref{coverage provability fixed n GC closed-form 1}, which is given on the next page.
\begin{figure*}[!t]
\normalsize
\begin{align}\label{coverage provability fixed n GC closed-form 1}
P_{{\rm{cov,up}}}^{k,ap}\left( {{\gamma _{th}}} \right) &\approx {\omega _M}\sum\limits_{m = 1}^M {{\vartheta _m}\frac{{{n}!{l_m}^{2k - 1}{{\left( {1 - {l_m}^2} \right)}^{{n} - k}}}}{{\left( {{n} - k} \right)!\left( {k - 1} \right)!}}} \nonumber  \\& \times \exp \left( { - {\rho _m}{\sigma ^2}} \right) {\left[ {{\omega _T}\sum\limits_{t = 1}^T {\frac{{{\mu _t}c_t^{\alpha  + 1}}}{{c_t^\alpha  + {\gamma _{th}}{{\tilde P}_x}^{ - 1}}}} } \right]^{k - 1}} \times {\left[ {\frac{{{\omega _T}}}{{1 - {l_m}^2}}\left( {\sum\limits_{t = 1}^T {\frac{{{\mu _t}c_t^{\alpha  + 1}}}{{c_t^\alpha  + {l_m}^\alpha {{\tilde P}_x}^{ - 1}{\gamma _{th}}}}}  - {l_m}^2\sum\limits_{t = 1}^T {\frac{{{\mu _t}c_t^{\alpha  + 1}}}{{c_t^\alpha  + {\gamma _{th}}{{\tilde P}_x}^{ - 1}}}} } \right)} \right]^{n - k}}\nonumber\\
& \times \exp \left[ { - {a^2}l_m^2\pi {\lambda _G}\delta \sum\limits_{p = 1}^n {
n\choose
p} B\left( {p - \delta ,n - p + \delta } \right){{\left( {{\gamma _{th}}{{\tilde P}_x}^{ - 1}} \right)}^\delta }} \right. \left. { - \frac{{{a^2}l_m^2{\pi ^2}\delta }}{{\sin \left( {\pi \delta } \right)}}{\lambda _{{\rm{co}}}}{{\left( {{\gamma _{th}}{{\tilde P}_z}^{ - 1}} \right)}^\delta }} \right].
\end{align}
\hrulefill \vspace*{0pt}
\end{figure*}

For the cluster  with a random number of active LoRa nodes of average $\bar n$, the coverage probability of the $k$-th closest LoRa node can be approximated as
\begin{figure*}[!t]
\normalsize
\begin{align}\label{coverage provability worst mean n GC closed-form 2}
P_{{\rm{cov,low}}}^{w,app}\left( {{\gamma _{th}}} \right) &\approx {\omega _M}\sum\limits_{m = 1}^M {{\vartheta _m}\frac{{\bar n!{l_m}^{2k - 1}{{\left( {1 - {l_m}^2} \right)}^{\bar n - k}}}}{{\left( {\bar n - k} \right)!\left( {k - 1} \right)!}}}  \times \exp \left( { - {\rho _m}{\sigma ^2}} \right)\exp \left( { - \left( {\bar n - 1} \right){\omega _T}\sum\limits_{t = 1}^T {\frac{{{\mu _t}{c_t}}}{{c_t^\alpha {{\tilde P}_x}{{\left( {{\gamma _{th}}} \right)}^{ - 1}} + 1}}} } \right)\nonumber\\
&\times\exp \left[ { - \frac{{{a^2}l_m^2{\pi ^2}\delta }}{{\sin \left( {\pi \delta } \right)}}\left( {{\lambda _G}\bar n{{\left( {{\gamma _{th}}{{\tilde P}_x}^{ - 1}} \right)}^\delta } + {\lambda _{{\rm{co}}}}{{\left( {{\gamma _{th}}{{\tilde P}_z}^{ - 1}} \right)}^\delta }} \right)} \right].
\end{align}
\hrulefill \vspace*{0pt}
\end{figure*}

\end{corollary}

\section{Area Spectral Efficiency and Energy Efficiency}
Spectral efficiency and energy efficiency are two critical factors for the design of LPWA networks, as LPWA networks  are required to support massive connectivity with minimal energy consumption at sensor nodes. We will investigate the area spectral efficiency and energy efficiency in this section.
\subsection{Area Spectral Efficiency}
The area spectral efficiency is defined as the average data in bits that all transmitters can contribute per unit area. Therefore, based on section III, we can get the area spectral efficiency in the following proposition.

\begin{proposition}\label{Network Spectrum Efficiency proposition}
The area spectral efficiency of a given network can be expressed as
\begin{align}\label{Network Spectrum Efficiency}
{\tau  } = {n_\iota }{\lambda _G}{R_t}P_{{\mathop{\rm cov}} }^{ap},
\end{align}
where ${R_t} = {\log _2}\left( {1 + {\gamma _{th}}} \right)$ is the transmission rates of all LoRa nodes, $P_{{\mathop{\rm cov}} }^{ap}$ is given by \eqref{coverage provability fixed closed-form 1}, \eqref{coverage provability closed-form 1}, \eqref{coverage provability fixed n GC closed-form 1}, and \eqref{coverage provability worst mean n GC closed-form 2} for the four scenarios considered in this paper, respectively.
\end{proposition}

If the density of LoRa receivers $\lambda_G$, the density of non-LoRa receivers $\lambda_{\rm co}$, and the number of active LoRa nodes in each cluster ${n_\iota } = \left\{ {n,\bar n} \right\}$ are fixed, according to~\eqref{Network Spectrum Efficiency}, we have the following remarks.

\begin{remark}\label{spectrum efficiency closed-form 2}
The area spectral efficiency  is a monotonically decreasing function of the cluster radius $a$ according to~\eqref{Network Spectrum Efficiency}. Therefore, the area spectral efficiency can be enhanced by densely deploying LoRa nodes around each LoRa receiver.
\end{remark}

\begin{remark}\label{spectrum efficiency_number_of_LoRa_nodes}
The area spectral efficiency  is not a monotonic function of the number of active LoRa nodes $n_{\iota}$ in the cluster according to~\eqref{Network Spectrum Efficiency}. Therefore, the area spectral efficiency can be maximized by adjusting the number of active LoRa nodes around each LoRa receiver properly. Due to the complex expressions of \eqref{coverage provability fixed closed-form 1}, \eqref{coverage provability closed-form 1}, \eqref{coverage provability fixed n GC closed-form 1}, and \eqref{coverage provability worst mean n GC closed-form 2}, the number of active LoRa nodes in the cluster that achieves the maximal area spectral efficiency is shown in the simulation results.
\end{remark}

\subsection{Energy Efficiency}
For a given network, the total power consumption for  nodes within per unit area is ${n_\iota }{\lambda _G}{P_{x}}$, where $P_x$ is the transmit power. Energy efficiency is defined as the ratio of the total amount of data delivered and the total energy consumed.
\begin{proposition}
Based on \eqref{Network Spectrum Efficiency}, we can obtain the energy efficiency of the considered network as
\begin{align}\label{Network Energy Efficiency}
\mathrm{EE} = \frac{{{n_\iota }{\lambda _G}{R_t}P_{{\rm{cov}}}^{ap}}}{{{n_\iota }{\lambda _G}{P_{x}}}} = \frac{{{R_t}P_{{\rm{cov}}}^{ap}}}{{{P_{x}}}},
\end{align}
where $R_t$ and $P_{{\rm{cov}}}^{ap}$ are same as defined in~\eqref{Network Spectrum Efficiency}.

\end{proposition}
\begin{remark}\label{Network Energy Efficiency 1}
Energy efficiency  is a monotonically decreasing function of cluster radius, $a$. Therefore, the energy efficiency of a given network can be enhanced by densely deploying LoRa nodes around each LoRa receiver.
\end{remark}

\section{Numerical Results}
We first validate the analytical results presented in the earlier sections and investigate the tightness of various approximations derived for coverage probability. In our simulation, the locations of LoRa nodes are drawn from a PCP over a circle region with radius $R_n=20$~km as LoRa claims that it is able to support long range transmission in terms of kilometers. The simulation parameters are set according to the LoRa specifications unless stated otherwise. The transmission frequency is $f_c=868$ MHz as it is used for LoRa in Europe~\cite{LoRa_specification}. The bandwidth is $BW=125$ KHz as one of most common setting-ups for LoRa networks. The thermal noise in dBm level is calculated as ${\sigma ^2} =  - 174 + 10{\log _{10}}\left( {BW} \right)$. The path-loss exponent for the communication links is $\alpha=3.5$. For the ordered case, we take $k=\mathcal{N}$ to show the achievable performance of the LoRa node that locates farthest away from the LoRa receiver.

\begin{figure}[t!]
    \begin{center}
        \includegraphics[width= 3.4in, height=2.3in]{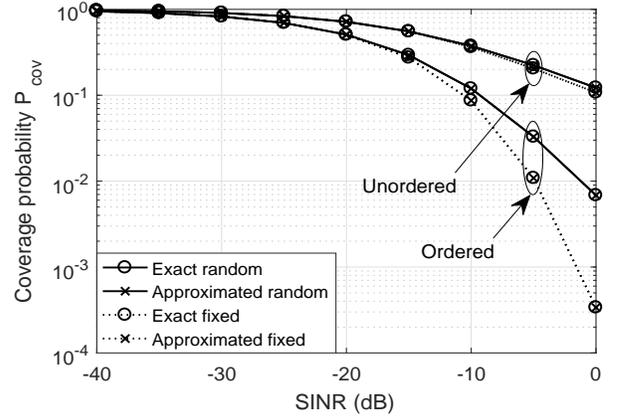}
        \caption{Verification of exact and approximated results versus SINR thresholds $\gamma_{th}$, transmit power is $P_{x}=14$~dBm, and density of LoRa receivers and coexisted non-LoRa nodes are ${\lambda _G}={\lambda _{\rm co}} = {10^{-1}}/\left( {{{500}^2}\pi } \right)$, cluster radius is $a=500$~m, and the number of active LoRa nodes in the cluster is $n_\iota =6$.}
        \label{interger_approx}
    \end{center}
\end{figure}

Fig.~\ref{interger_approx} plots the coverage probability of the networks with different SINR thresholds at the receivers $\gamma_{th}$. We can see that  the derived closed-form expressions in Corollaries~\ref{coverage provability fixed closed-form} and~\ref{coverage provability fixed n GC closed-form} are well matched with Theorems~\ref{coverage provability exact} and ~\ref{coverage provability k nearest ordered}. Therefore, we use the approximations given in Corollaries~\ref{coverage provability fixed closed-form} and~\ref{coverage provability fixed n GC closed-form} as the analytical results in the following figures. It is also worth noting that the unordered case  always achieves better coverage probability than the ordered case. The reason is that in the ordered case  the farthest LoRa node is chosen as the the typical one by taking $k=\mathcal{N}$. For the unordered case, the LoRa node of interest is selected randomly among the active nodes in the same cluster, which results in an average performance of the considered networks.

\begin{figure*}[t!]
    \centering
    \subfigure[]{\label{sim_ana_fixed}
\includegraphics[width= 3.2in, height=2.2in]{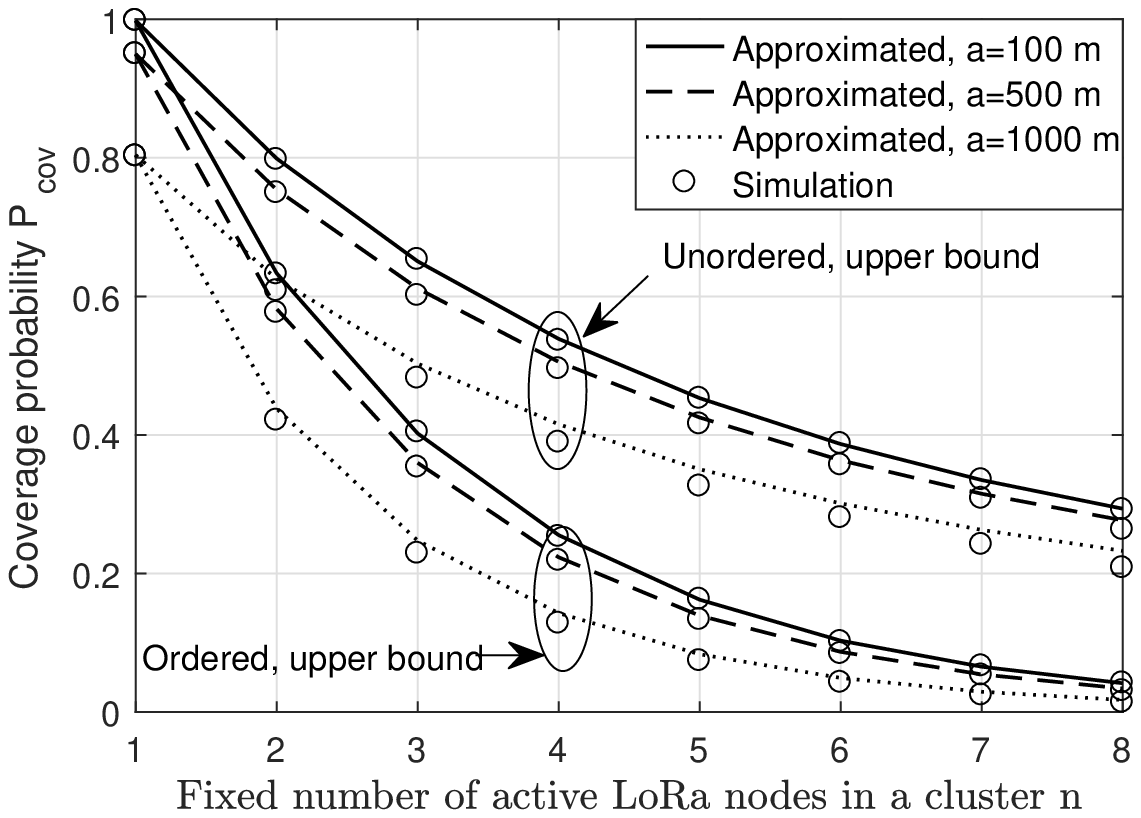}}
    \subfigure[]{\label{sim_ana_mean}
\includegraphics[width= 3.2in, height=2.2in]{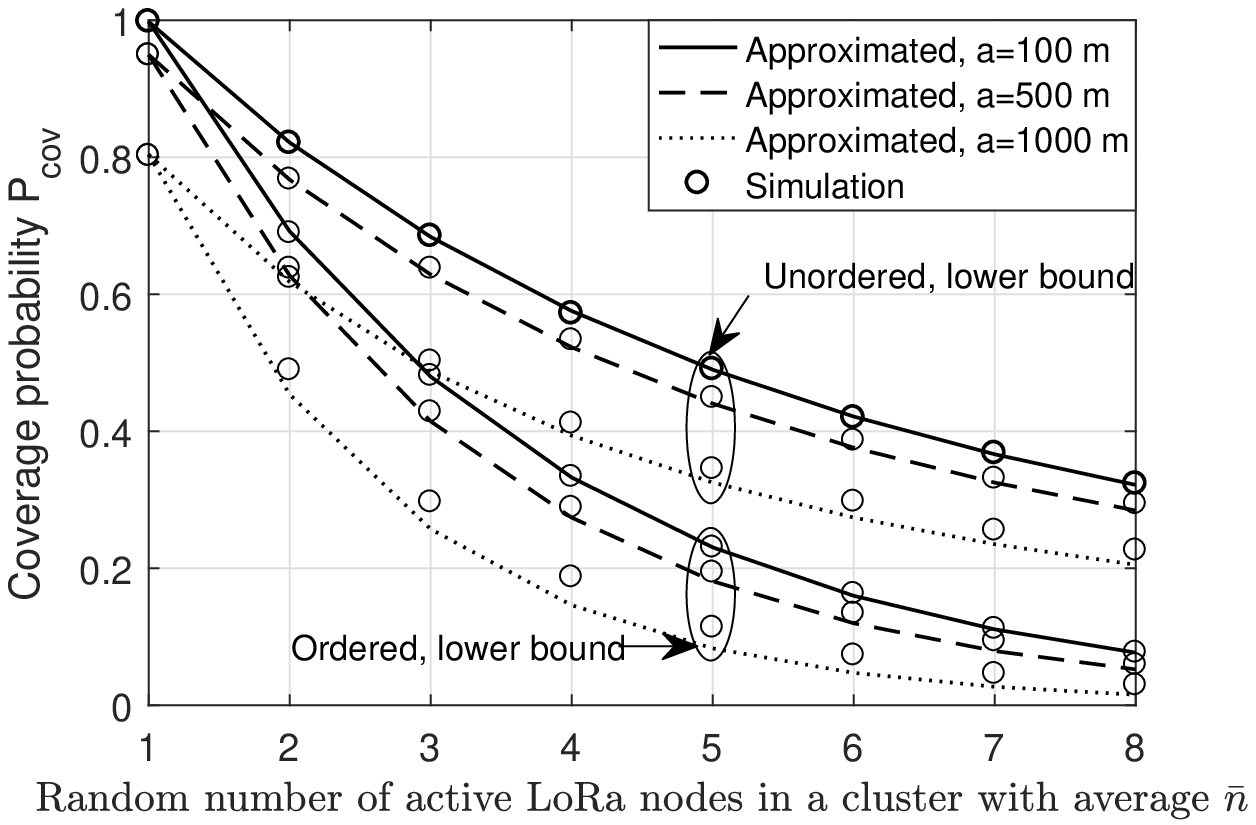}}
    \caption{Verification of approximated and simulation results versus number of active LoRa nodes in each cluster with different cluster radii $a$, density of LoRa receivers and non-LoRa nodes are $\lambda_{G}=\lambda_{\rm co}={10^{-1}}/\left( {{{500}^2}\pi } \right)$, transmit power is $P_{x}=14$~dBm, SINR threshold is $\gamma_{th}=-10$~dB.}
    \label{approx_simulation}
\end{figure*}

Fig.~\ref{approx_simulation} plots the coverage probability of the networks for the ordered and unordered cases versus the number of active LoRa nodes in each cluster $n_\iota$. For the curve, we can observe the following:
\begin{enumerate}
\item The coverage probability decreases with larger cluster radius $a$, which is consistent with the discussion in Remark \ref{coverage provability remark 2}. This is due to the fact that larger $a$ not only increases the distance of the desired link but also the inter-cluster interference, because larger $a$ shortens  the distance between sources of inter-cluster interference and the typical receiver.
\item It is also worth noting that the approximated results given in~\eqref{coverage provability fixed closed-form 1},~\eqref{coverage provability closed-form 1},~\eqref{coverage provability fixed n GC closed-form 1} and~\eqref{coverage provability worst mean n GC closed-form 2} are well matched with the simulation results when the cluster radius is $a=100$~m. With increasing $a$, there is a small but notable gap between the approximated results and the simulation results, which is caused by  using the distance approximation, $q\left( {x,y,\theta } \right) \approx y$, in Section III. In a practical scenario of LPWA networks, the cluster radius $a$ is typically a value in terms of kilometers. Here, we provide results with $a=100$~m is just to justify the tightness of our derived approximated results.
\end{enumerate}

As observed from Fig.~\ref{interger_approx} and Fig.~\ref{approx_simulation},  the unordered case behaves similarly to the ordered case. In order to save space without losing any important insights, in the following figures, we only present the results of the ordered case with a fixed and a random number of active LoRa nodes, respectively.

\begin{figure}[t!]
    \begin{center}
        \includegraphics[width=3.4in,height=2.3in]{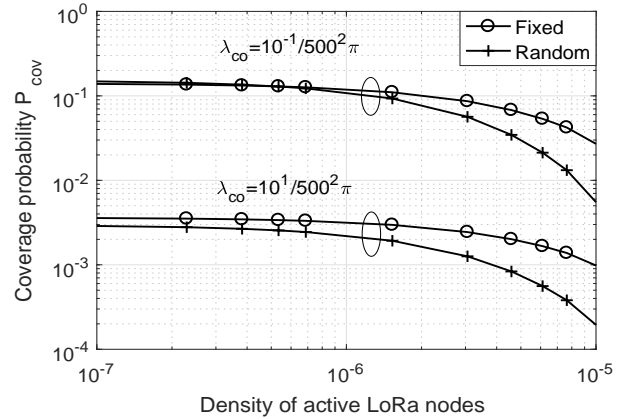}
        \caption{Coverage probability versus density of ordered active LoRa nodes with different density of non-LoRa nodes $\lambda_{\rm co}$, transmit power is $P_{x}=14$~dBm, SINR threshold is $\gamma_{th}=-10$~dB, cluster radius is $a=500$~m, number of active LoRa nodes in the cluster is $n_{\iota} =6$.}
        \label{different_lambda_3D}
    \end{center}
\end{figure}

Fig.~\ref{different_lambda_3D} plots the coverage probability of the networks for ordered cases versus density of active LoRa nodes ${n_\iota }\lambda_G$ with different densities of non-LoRa nodes $\lambda_{\rm co}$. Both the cases with fixed and random numbers of active LoRa nodes in each cluster are presented. It is worth noting that coverage probability is monotonically decreasing with density of active LoRa nodes ${n_\iota }\lambda_G$ and the density of non-LoRa nodes $\lambda_{\rm co}$. The reason is that larger densities of active LoRa nodes and non-LoRa nodes bring higher intra-interference, inter-interference, and coexistence interference. This phenomenon is consistent with the discussion in Remark~\ref{coverage provability remark 1}.

\begin{figure}[t!]
    \begin{center}
    \includegraphics[width= 3.4in, height=2.3in]{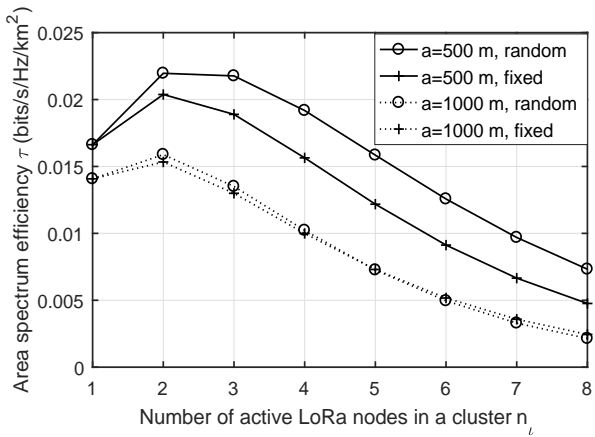}
    \caption{Area spectral efficiency versus the number of ordered active LoRa nodes in each cluster for ordered and unordered cases with different cluster radii $a$, density of LoRa receivers and non-LoRa nodes are ${\lambda _G} = {\lambda _{\rm co}}={10^{-1}}/\left( {{{500}^2}\pi } \right)$, transmit power is $P_{x}=14$~dBm, and SINR threshold is $\gamma_{th}=-10$~dB.}
    \label{ASE}
    \end{center}
\end{figure}

Fig.~\ref{ASE} plots the area spectral efficiency of the networks versus the number of active LoRa nodes $n_\iota$ simultaneously transmitting in each cluster with different cluster radii $a$. From the curve, the area spectral efficiency is monotonically decreasing with the cluster radius $a$, which is consistent with the discussion in Remark \ref{spectrum efficiency closed-form 2}. We can also observe that the area spectral efficiency is not a monotonic function of ${n_\iota } $ as discussed in Remark \ref{spectrum efficiency_number_of_LoRa_nodes}. In other words, there exists an optimal number of active LoRa links. This behavior can be explained as follows: on the one hand, more simultaneously transmitting LoRa links bing larger intra-cluster interference, as such, the coverage probability decreases, which in turn decreases the area spectral efficiency. On the other hand, as seen from \eqref{Network Spectrum Efficiency}, larger ${n_\iota } $ results in more efficient spectrum unitization per unit area, which enhances the area spectral efficiency.

\begin{figure}[t!]
\begin{center}
\includegraphics[width= 3.4in, height=2.3in]{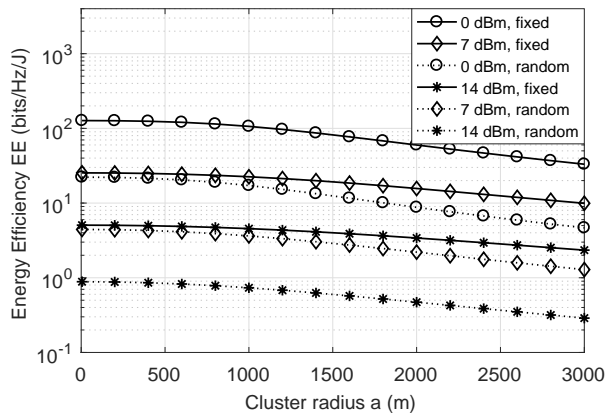}
    \caption{Energy efficiency versus cluster radius $a$ with different transmit powers $P_{x}$, density of LoRa receivers and non-LoRa nodes  are ${\lambda _G} = {\lambda _{\rm co}}={10^{-1}}/\left( {{{500}^2}\pi } \right)$, and SINR threshold is $\gamma_{th}=-10$~dB.}
    \label{EE}
    \end{center}
\end{figure}

Fig.~\ref{EE} plots the energy efficiency of the networks versus the cluster radius $a$ with different transmit powers $P_x$. We can see that the energy efficiency is monotonically decreasing with the cluster radius $a$, which is consistent with the discussion in Remark \ref{Network Energy Efficiency 1}. We also observe that the energy efficiency is not a monotonically decreasing function with the transmit power $P_x$ according to~\eqref{Network Energy Efficiency}. However, when the transmit power is set to $0$~dBm, $7$~dBm and $14$~dBm as specified in LoRa networks, the energy efficiency decreases monotonically.

\begin{figure*}[t!]
    \centering
    \subfigure[]{\includegraphics[width= 3.2in, height=2.2in]{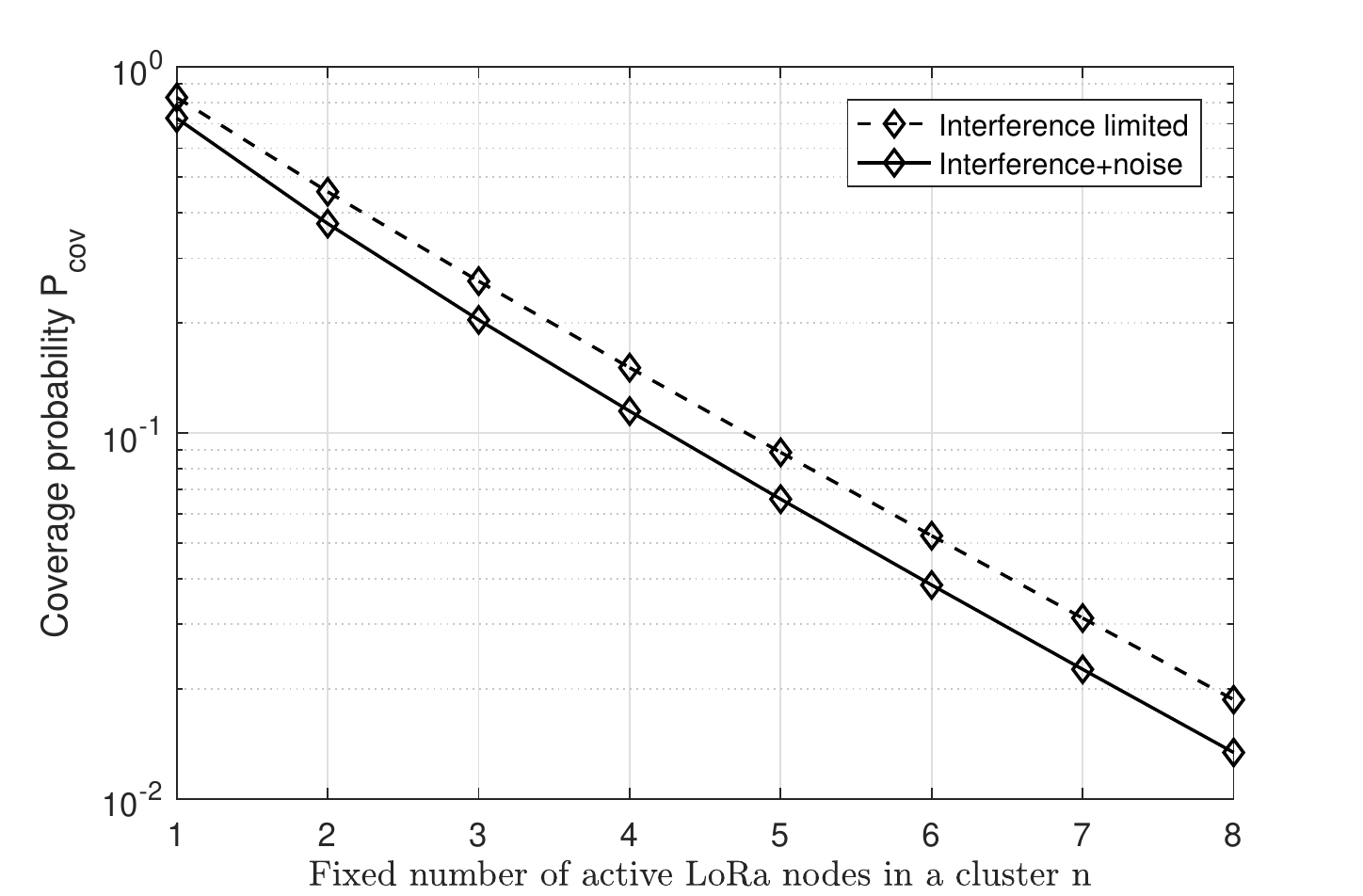}}
    \subfigure[]{\includegraphics[width= 3.2in, height=2.2in]{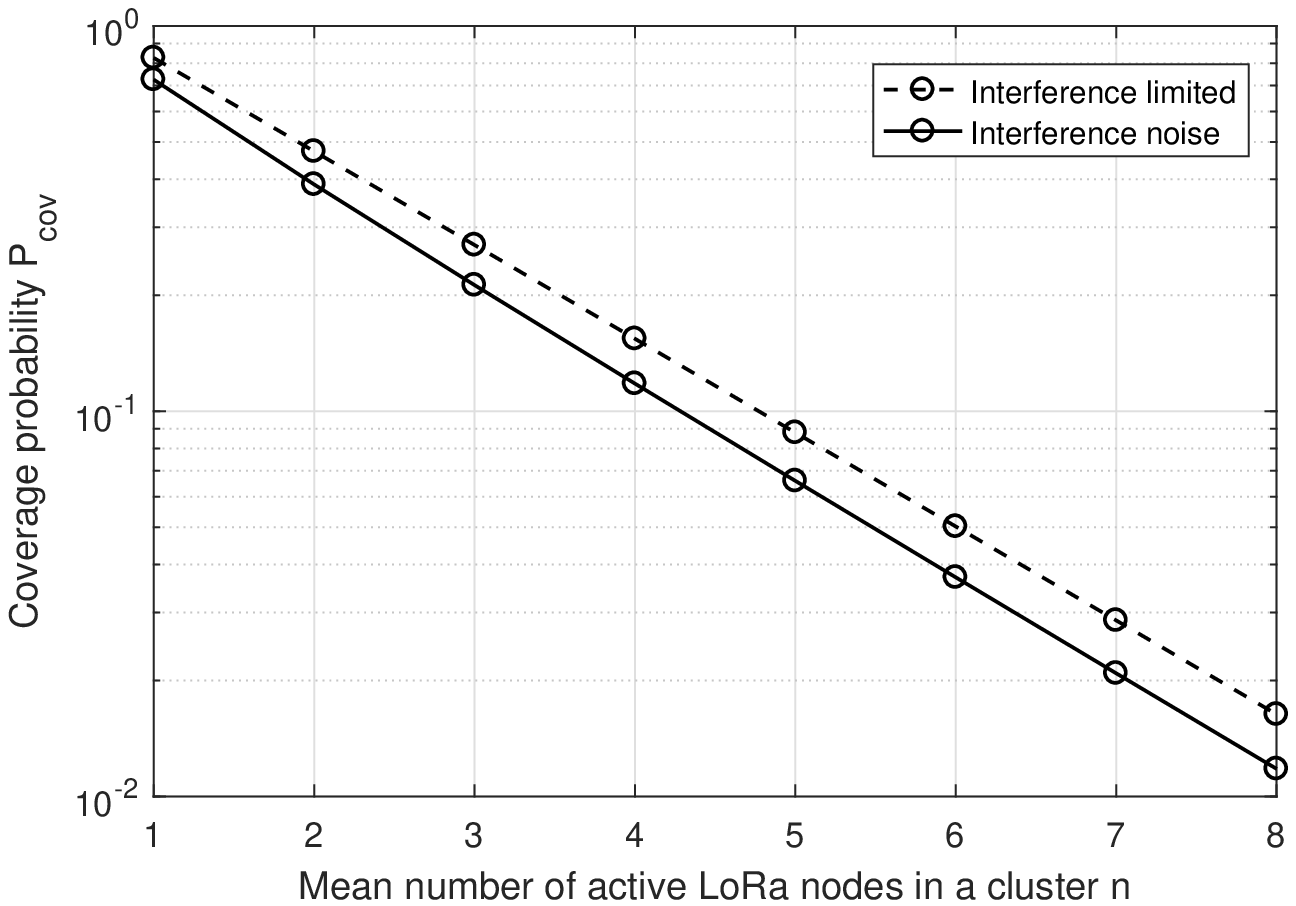}}
    \caption{Coverage probability versus the number of active LoRa nodes in each cluster, density of LoRa receivers and non-LoRa nodes are $\lambda_{G}=\lambda_{\rm co}={10^{-1}}/\left( {{{500}^2}\pi } \right)$, cluster radius is $a=1000$~m, transmit power is $P_{x}=7$~dBm, SINR threshold is $\gamma_{th}=-10$~dB.}
    \label{interference_limited}
\end{figure*}

Fig.~\ref{interference_limited} verifies the necessity  of considering noise in the analysis by comparing the coverage probability for cases with and without noise. It can be shown that the coverage probability of the interference-limited case is always higher than that with both interference and noise. This is mainly because that the distance from LoRa users to gateway is usually very long with low transmit power in LPWA networks.

As part of a smart city project, the above observations are capable of providing insights when determining the system parameters of LoRa networks in practical scenarios. For example, with the purpose of maximizing the area spectral efficiency of an LPWA network, there exists an optimal number of LoRa nodes transmitting to the the LoRa receiver simultaneously. This inspires the strategy for when and how to scale the LoRa network by introducing new LoRa gateways in order to achieve the optimal area spectral efficiency. Another example is that in order to improve the energy efficiency of an LPWA network, transmit power should be minimized if the LoRa node is within the coverage of a LoRa receiver.

\section{Conclusions}
In this paper, the uplink transmission of low-power wide-area (LPWA)  networks  with multiple radio modules has been studied. By using LoRa as an application of our technique and  accounting for coexistence interference from other types of radio modules sharing the space, we investigated the performance of the LPWA networks with invoking the poisson cluster process (PCP) to model the locations of LoRa nodes. Specifically, we consider the scenarios that the number of active LoRa nodes in each cluster is fixed and random, respectively, and that the typical LoRa node is chosen randomly or according to the distance from the typical node to its serving receiver.  Besides the exact expressions, we have also derived the simple and approximated expressions for the coverage probability of a typical LoRa node, its area spectral efficiency, and its energy efficiency. According to our analyses,  an effective approach for enhancing network performance is to deploy nodes more densely but there exists an optimal number of active LoRa nodes in each cluster to maximize the area spectral efficiency. Furthermore, energy efficiency decreases monotonically with the transmit power using the values that are specified in LoRa specification. As part of a smart city project, our results here can provide insightful guidelines to inform our real world deployment of the large-scale LPWA networks.

\section{Acknowledgement}

The authors would like to thank Dr. Mo Haghighi from Intel Labs Europe for his clarification regarding the specifics \& LoRa protocol.

\numberwithin{equation}{section}
\section*{Appendix~A: Proof of Lemma~\ref{Intra Laplace unordered fixed n}} \label{Appendix:A}
\renewcommand{\theequation}{A.\arabic{equation}}
For the unordered case with $n$ active LoRa nodes in each cluster, the Laplace transforms of intr-cluster interference is given by
\setcounter{equation}{0}
\begin{align}\label{Intra Laplace unordered fixed n proof 1}
&{{\cal L}_{I_{{\rm{intra}}}^n}}\left( s \right) =  {\mathbb{E}_{x \in {N_{y_0}}\backslash {x_0}}}\left[ {\exp \left( { - s\sum\limits_{x \in {N}\backslash {x_0}} {{P_{x}}{h_{x,y_0}}\eta {{\left\| x \right\|}^{ - \alpha }}} } \right)} \right]\nonumber\\
&= {\mathbb{E}_{x \in N_{y_0}\backslash {x_0}}}\left[ {\prod\limits_{x \in {N_{y_0}}} {{\mathbb{E}_{{h_{x,y_0}}}}\left[ {\exp \left( { - s{P_{x}}{h_{x,y_0}}\eta {{\left\| x \right\|}^{ - \alpha }}} \right)} \right]} } \right]\nonumber\\
&\mathop  = \limits^{\left( a \right)} {\mathbb{E}_{x \in N_{y_0}\backslash {x_0}}}\left[ {\prod\limits_{x \in {N_{y_0}}\backslash {x_0}} {\frac{1}{{1 + s{P_{x}}\eta {{\left\| x \right\|}^{ - \alpha }}}}} } \right]\nonumber\\
&={\left[ {\int_R {\left( {\frac{1}{{1 + s{P_x}\eta {{\left\| x \right\|}^{ - \alpha }}}}} \right)} {f_{\left\| x \right\|}}\left( x \right)dx} \right]^{n - 1}}\nonumber\\
&\mathop  = \limits^{\left( b \right)} {\left[ {\int_0^a {\left( {\frac{1}{{1 + s{P_x}\eta {r^{ - \alpha }}}}} \right)} {f_R}\left( r \right)dr} \right]^{n - 1}},
\end{align}
where $(a)$ follows the fact that ${{h_{x,y_0}}}$ follows a Rayleigh fading distribution with unit mean, $(b)$ is obtained by changing the integration from the Cartesian coordinates into the polar coordinates. Plugging \eqref{PDF_polar} into \eqref{Intra Laplace unordered fixed n proof 1} and letting $t = {r^\alpha }$, we can obtain
\begin{align}\label{Intra Laplace unordered fixed n proof 2}
{{\cal L}_{I_{{\rm{intra}}}^n}}\left( s \right) = {\left[ {\frac{\delta }{{{a^2}s{P_x}\eta }}\int_0^{{a^\alpha }} {\left( {\frac{{{t^\delta }}}{{t{{\left( {s{P_x}\eta } \right)}^{ - 1}} + 1}}} \right)} dt} \right]^{n - 1}}.
\end{align}
Then applying \cite[Eq. (3.194).1]{gradshteyn}, we can obtain \eqref{Intra Laplace unordered fixed n 1}.

Following the similar procedure of obtaining \eqref{Intra Laplace unordered fixed n 1},  we have
\begin{align}\label{Intra Laplace proof 1}
&{{\cal L}_{I_{{\rm{intra}}}^{\bar n}}}\left( s \right)
\mathop  = \limits^{\left( a \right)} {\mathbb{E}_{x \in N_{y_0}\backslash {x_0}}}\left[ {\prod\limits_{x \in {N_{y_0}}\backslash {x_0}} {\frac{1}{{1 + s{P_{x}}\eta {{\left\| x \right\|}^{ - \alpha }}}}} } \right]\nonumber\\
&\mathop  = \limits^{\left( b \right)} \exp \left[ {\left( {\bar n - 1} \right)\int_{R^+} {\left( {1 - \frac{1}{{1 + s{P_x}\eta {{\left\| x \right\|}^{ - \alpha }}}}} \right)} {f_{\left\| x \right\|}}\left( x \right)dx} \right],
\end{align}
where $(a)$ follows the fact that ${{h_{x,y_0}}}$ follows a Rayleigh fading distribution with unit mean, $(b)$ is obtained by applying the moment-generating function. Substituting \eqref{PDF_Rayleigh} into \eqref{Intra Laplace proof 1} and changing the integration from the Cartesian coordinates into the polar, we have
\begin{align}\label{Intra Laplace 2}
{{\cal L}_{I_{{\rm{intra}}}^{\bar n}}}\left( s \right) = \exp \left( { - \frac{{2\left( {\bar n - 1} \right)}}{{{a^2}}}\int_0^a {\frac{{s{P_x}\eta {r^{ - \alpha }}}}{{1 + s{P_x}\eta {r^{ - \alpha }}}}} rdr} \right).
\end{align}
With the aid of \cite[Eq. (3.194).1]{gradshteyn}, we obtain \eqref{Intra Laplace 1}. The proof is completed.


\section*{Appendix~B: Proof of~\eqref{Inter Laplace fixed lower bound_1} in Lemma~\ref{Inter Laplace fixed lower bound}} \label{Appendix:B}
\renewcommand{\theequation}{B.\arabic{equation}}
\setcounter{equation}{0}
Based on \eqref{I_inter}, the the Laplace transform for inter-cluster interference with  $n$ active LoRa nodes in the cluster is given by
\begin{align}\label{Inter Laplace fixed lower bound proof_1}
&{\mathcal{L}_{I_{{\rm{inter}}}^{n}}}\left( s \right)\nonumber\\& = {\mathbb{E}_{{\Phi _G}}}\left[ {\prod\limits_{y \in {\Phi _G}\backslash {y_0}} {\exp \left( { - s\sum\limits_{x \in {N_y}} {{P_x}{h_{x,y}}\eta {{\left\| {y + x} \right\|}^{ - \alpha }}} } \right)} } \right]\nonumber\\
&\mathop  = \limits^{\left( a \right)} {\mathbb{E}_{{\Phi _G}}}\left[ {\prod\limits_{y \in {\Phi _G}\backslash {y_0}} {{\mathbb{E}_x}\left[ {\prod\limits_{x \in {N_y}} {\left( {\frac{1}{{1 + s{P_x}\eta {{\left\| {y + x} \right\|}^{ - \alpha }}}}} \right)} } \right]} } \right]\nonumber\\
&\mathop {{\rm{  }} = }\limits^{\left( b \right)} \exp \left[ { - {\lambda _G}} \right. \times \nonumber\\& \left. {\int_{{R^2}} {\left( {1 - {{\left[ {{E_x}\left( {\frac{1}{{{{\left( {1 + s{P_x}\eta {{\left\| {y + x} \right\|}^{ - \alpha }}} \right)}^n}}}} \right)} \right]}^n}} \right)dy} } \right],
\end{align}
where $(a)$ is obtained by ${{h_{x,y}}}$ follows the Rayleigh distribution with unit mean, $(b)$ is obtained with using the generating functional of \emph{Matern} point process with a fixed number of points $n$ in each cluster.
With the aid of Jensen inequality  ${\left[ {{\mathbb{E}_x}\left( x \right)} \right]^n} \le {\mathbb{E}_x}\left( {{{\left( x \right)}^n}} \right)$, we can obtain a tight upper bound for the Laplace transform  of inter-cluster interference, which is given by

\begin{align}\label{Inter Laplace fixed lower bound proof_1}
&{\mathcal{L}_{I_{{\rm{inter}}}^{n}}}\left( s \right) \le \mathcal{L}_{I_{{\rm{inter}}}^{n}}^{up}\left( s \right) \nonumber\\
&\mathop {{\rm{  }} = }\limits^{\left( a \right)} \exp \left[ { - {\lambda _G}\int_{{R^2}} {{f_{\left\| x \right\|}}\left( x \right)dx} } \right.)\nonumber\\& \times \left. {\left( {\int_{{R^2}} {\left[ {1 - \frac{1}{{{{\left( {1 + s{P_x}\eta {{\left\| {{y^*}} \right\|}^{ - \alpha }}} \right)}^n}}}} \right]d} {y^*}} \right)} \right]\nonumber\\
&\mathop  = \limits^{\left( b \right)} \exp \left( { - 2\pi {\lambda _G}\int_0^\infty  {\left( {1 - \frac{1}{{{{\left( {1 + s{P_x}\eta {r^{ - \alpha }}} \right)}^n}}}} \right)} rdr} \right)\nonumber\\
&\mathop  = \limits^{\left( c \right)} \exp \left( { - 2\pi {\lambda _G}\sum\limits_{p = 1}^n {
n\choose
p} {{\left( {s{P_x}\eta } \right)}^p}\int_0^\infty  {\frac{{{r^{ - \alpha p + 1}}}}{{{{\left( {1 + s{P_x}\eta {r^{ - \alpha }}} \right)}^n}}}} dr} \right)\nonumber\\
&\mathop  = \limits^{\left( d \right)} \exp \left( { - \pi {\lambda _G}\sum\limits_{p = 1}^n {
n\choose p
} {{\left( {s{P_x}\eta } \right)}^\delta }\delta \int_0^\infty  {\frac{{{{\left( t \right)}^{p - \delta  - 1}}}}{{{{\left( {1 + t} \right)}^n}}}} dt} \right)\nonumber\\
&\mathop  = \limits^{\left( e \right)} \exp \left( { - \pi {\lambda _G}{{\left( {s{P_x}\eta } \right)}^\delta }\delta \sum\limits_{p = 1}^n {
n\choose
p} B\left( {p - \delta ,n - p + \delta } \right)} \right),
\end{align}
where $(a)$ is obtained by changing variables as ${y^*} \to y + x$, $(b)$ is obtained by changing from cartesian to polar coordinate, $(c)$ is obtained by applying  Binomial expansion, $(d)$ is resulted from using the variable changes of $t = \frac{{s{P_x}\eta }}{{{r^\alpha }}}$, and $(e)$ is simplified with applying the definition of Beta function \cite[ Eq. (8.380) .3]{gradshteyn}.

\section*{Appendix~C: Proof of~\eqref{Inter Laplace final_appro_align} in Lemma~\ref{Inter Laplace fixed lower bound}} \label{Appendix:C}
\renewcommand{\theequation}{C.\arabic{equation}}
\setcounter{equation}{0}
Similar to \eqref{Inter Laplace fixed lower bound proof_1}, the the Laplace transform for inter-cluster interference with a random number of active LoRa nodes in the cluster with an average $\bar n$ is written by
\begin{align}\label{Inter Laplace Matern proof}
&{\mathcal{L}_{I_{{\rm{inter}}}^{\bar n}}}\left( s \right)\nonumber\\&={\mathbb{E}_{{\Phi _G}}}\left[ {\prod\limits_{y \in {\Phi _G}\backslash {y_0}} {{\mathbb{E}_x}\left[ {\prod\limits_{x \in {N_y}} {\left( {\frac{1}{{1 + s{P_x}\eta {{\left\| {y + x} \right\|}^{ - \alpha }}}}} \right)} } \right]} } \right].
\end{align}
Applying the the generating functional of \emph{Matern} point process with a random number of points of average $\bar n$, and similar to \eqref{Inter Laplace fixed lower bound proof_1} with the variables changes of ${y^*} \to y + x$, we can obtain
\begin{align}\label{Inter Laplace Matern proof 1}
&{\mathcal{L}_{I_{{\rm{inter}}}^{\bar n}}}\left( s \right)\nonumber\\ & =\exp \left( { - {\lambda _G}\int_{{R^2}} {\left( {1 - \exp \left( { - \bar n\left( {\frac{{s{P_x}\eta {y^*}^{ - \alpha }}}{{1 + s{P_x}\eta {y^*}^{ - \alpha }}}} \right)} \right)} \right)} d{y^*}} \right)\nonumber\\
&= \exp \left( { - 2\pi {\lambda _G}\int_0^\infty  {\left[ {1 - \exp \left( { - \bar n\left( {\frac{{s{P_x}\eta {r^{ - \alpha }}}}{{1 + s{P_x}\eta {r^{ - \alpha }}}}} \right)} \right)} \right]r} dr} \right).
\end{align}


With the aid of Taylor series expansion, we can have the approximation as $1 - \exp \left( { - x} \right) \le x$. Then we can obtain
\begin{align}\label{Inter Laplace approx Matern 1}
{{\cal L}_{I_{{\rm{inter}}}^{\bar n}}}\left( s \right) &\ge {\cal L}_{I_{{\rm{inter}}}^{\bar n}}^{low}\left( s \right)\nonumber\\
 &= \exp \left( { - 2\pi {\lambda _G}\bar n\int_0^\infty  {\left( {\frac{{s{P_x}\eta {r^{ - \alpha }}}}{{1 + s{P_x}\eta {r^{ - \alpha }}}}} \right)r} dr} \right)\nonumber\\
&\mathop  = \limits^{\left( a \right)} \exp \left( { - \pi {\lambda _G}\bar n {{\left( {s{P_{x}}\eta } \right)}^\delta }\delta\Gamma \left( \delta  \right)\Gamma \left( {1 - \delta } \right)} \right)\nonumber\\
 &\mathop  = \limits^{\left( b \right)}  \exp \left( { - {\pi ^2}{\lambda _G}\bar n{{\left( {s{P_{x}}\eta } \right)}^\delta }\frac{\delta }{{\sin \left( {\pi \delta } \right)}}} \right),
\end{align}
where $(a)$ is obtained by applying \cite[ Eq. (3.241).4]{gradshteyn}, and $(b)$ is obtained with the aid of the Euler's reflection formula.  The proof is completed.

\section*{Appendix~D: Proof of Lemma~\ref{Intra Laplace ordered fixed number}} \label{Appendix:D}
\renewcommand{\theequation}{D.\arabic{equation}}
\setcounter{equation}{0}
The Laplace transform of intra-cluster interference can be divided into two disjoint sets, as mentioned in Section III. By doing so, the Laplace transform of intra-cluster interference for the case with  $n$ active LoRa nodes in the cluster can be expressed as
\begin{align}\label{Intra Laplace ordered proof 1}
&{{\cal L}_{I_{{\rm{intra}}}^{n,k}}}\left( s \right)\nonumber\\&= \mathbb{E}\left[ {\prod\limits_{x \in {K_{near}}} {\frac{1}{{1 + s{P_x}\eta {{\left\| x \right\|}^{ - \alpha }}}}\prod\limits_{x \in {K_{far}}} {\frac{1}{{1 + s{P_x}\eta {{\left\| x \right\|}^{ - \alpha }}}}} } } \right]\nonumber\\
&\mathop {{\rm{  }} = }\limits^{\left( a \right)} {\left[ {\int_{{R^ + }} {\left( {\frac{1}{{1 + s{P_x}\eta {r^{ - \alpha }}}}} \right)} {f_{\left. {\tilde R} \right|r \le {{\tilde r}_k}}}\left( {\left. r \right|{{\tilde r}_k}} \right)dr} \right]^{k - 1}}  \nonumber\\
&\times{\left[ {\int_{{R^ + }} {\left( {1 - \frac{1}{{1 + s{P_x}\eta {r^{ - \alpha }}}}} \right)} {f_{\left. {\tilde R} \right|r > {{\tilde r}_k}}}\left( {\left. r \right|{{\tilde r}_k}} \right)dr} \right]^{n - k}}\nonumber\\
&\mathop {{\rm{  }} = }\limits^{\left( b \right)} {\left[ {\frac{2}{{{{\tilde r}_k}^2}}\underbrace {\int_0^{{{\tilde r}_k}} {\left( {\frac{1}{{1 + s{P_x}\eta {r^{ - \alpha }}}}} \right)rdr} }_{{Q_1}}} \right]^{k - 1}} \nonumber\\ &\times{\left[ {\frac{2}{{{a^2} - {{\tilde r}_k}^2}}\underbrace {\int_{{{\tilde r}_k}}^a {\left( {\frac{1}{{1 + s{P_x}\eta {r^{ - \alpha }}}}} \right)rdr} }_{{Q_2}}} \right]^{n - k}},
\end{align}
where $(a)$ is obtained by applying generating functional of \emph{Matern} point process with fixed number of points $n$, $(b)$ is obtained by plugging \eqref{Order PDF intra interfering 1} inside, which is PDF of the ordered distance for intra-cluster interference nodes.

Using the similar approach as obtaining \eqref{Intra Laplace unordered fixed n proof 2}, we can obtain
\begin{align}\label{Intra Laplace ordered Q1}
{Q_1} = \frac{\delta }{{2s{P_x}\eta }}\frac{{{{\tilde r}_k}^{\alpha  + 2}}}{{\delta  + 1}}{}_2{F_1}\left( {1,\delta  + 1;\delta  + 2; - \frac{{{{\tilde r}_k}^\alpha }}{{s{P_x}\eta }}} \right)
\end{align}
and
\begin{align}\label{Intra Laplace ordered Q2}
{Q_2} =& \frac{\delta }{{2s{P_x}\eta }}\frac{{{a^{\alpha  + 2}}}}{{\delta  + 1}}{F_1}\left( {1,\delta  + 1;\delta  + 2; - \frac{{{a^\alpha }}}{{s{P_x}\eta }}} \right)\nonumber\\&- \frac{\delta }{{2s{P_x}\eta }}\frac{{{{\tilde r}_k}^{\alpha  + 2}}}{{\delta  + 1}}{F_1}\left( {1,\delta  + 1;\delta  + 2; - \frac{{{{\tilde r}_k}^\alpha }}{{s{P_x}\eta }}} \right),
\end{align}
respectively. Substituting into \eqref{Intra Laplace ordered Q1} and \eqref{Intra Laplace ordered Q2} into \eqref{Intra Laplace ordered proof 1}, we can obtain \eqref{Intra Laplace ordered fixed number 1}.

The proof of~\eqref{Intra Laplace ordered worst 1} follows on the same procedure as~\eqref{Intra Laplace ordered fixed number 1} and is hence skipped.

\bibliographystyle{IEEEtran}
\bibliography{mybib}

\end{document}